# Probing Earth's Missing Potassium using the Unique Antimatter Signature of Geoneutrinos

A. Cabrera[*,12α,2,a], M. Chen[†,6], F. Mantovani[‡,3α,3β], A. Serafini[§,3α,3β,13α,13β], V. Strati[¶,3α,3β], J. Apilluelo[18], L. Asquith[1], J.L. Beney[11], T.J.C. Bezerra[1], M. Bongrand[11], C. Bourgeois[12α], D. Breton[12α], M. Briere[12α], J. Busto[10], A. Cadiou[11], E. Calvo[8], V. Chaumat[12α], E. Chauveau[4], B.J. Cattermole[1], P. Chimenti[7], C. Delafosse[12α], H. de Kerret[‖,a], S. Dusini[13α], A. Earle[1], C. Frigerio-Martins[7], J. Galán[18], J. A. García[18], R. Gazzini[12α], A. Gibson-Foster[1], A. Gallas[12α], C. Girard-Carillo[9α], W.C. Griffith[1], F. Haddad[11], J. Hartnell[1], A. Hourlier[17], G. Hull[12α], I. G. Irastorza[18], L. Koch[9α], P. Laniéce[12α,12β], J.F. Le Du[12α,2], C. Lefebvre[6], F. Lefevre[11], F. Legrand[12α], P. Loaiza[12α], J. A. Lock[1], G. Luzón[18], J. Maalmi[12α], C. Marquet[4], M. Martínez[18], B. Mathon[12α], L. Ménard[12α,12β], D. Navas-Nicolás[12α], H. Nunokawa[15], J.P. Ochoa-Ricoux[5], M. Obolensky[a], C. Palomares[8], P. Pillot[11], J.C.C. Porter[1], M.S. Pravikoff[4], H. Ramarijaona[12α], M. Roche[4], P. Rosier[12α], B. Roskovec[14], M.L. Sarsa[18], S. Schoppmann[9β], W. Shorrock[1], L. Simard[12α], H.Th.J. Steiger[9α,9β], D. Stocco[11], J.S. Stutzmann[11], F. Suekane[16,a], A. Tunc[9α], M.-A. Verdier[12α,12β], A. Verdugo[8], B. Viaud[11], S. M. Wakely[9α], A. Weber[9α], and F. Yermia[11]

(LiquidO Consortium)

[1] Department of Physics and Astronomy, University of Sussex, Brighton, United Kingdom
[2] LNCA Underground Laboratory, CNRS, EDF Chooz Nuclear Reactor, Chooz, France
[3α] INFN, Sezione di Ferrara, Ferrara, Italy
[3β] Dipartimento di Fisica e Scienze della Terra, Università di Ferrara, Ferrara, Italy
[4] Université de Bordeaux, CNRS, LP2I Bordeaux, Gradignan, France
[5] Department of Physics and Astronomy, University of California at Irvine, Irvine, CA, USA
[6] Department of Physics, Engineering Physics & Astronomy, Queen's University, Kingston, Canada
[7] Departamento de Física, Universidade Estadual de Londrina, Londrina, Brazil
[8] CIEMAT, Centro de Investigaciones Energéticas, Medioambientales y Tecnológicas, Madrid, Spain
[9α] Johannes Gutenberg-Universität Mainz, Institut für Physik, Mainz, Germany
[9β] Johannes Gutenberg-Universität Mainz, Detektorlabor, Exzellenzcluster PRISMA+, Mainz, Germany
[10] Université de Aix Marseille, CNRS, CPPM, Marseille, France
[11] Nantes Université, IMT-Atlantique, CNRS, Subatech, Nantes, France
[12α] Université Paris-Saclay, CNRS/IN2P3, IJCLab, Orsay, France
[12β] Université Paris Cité, IJCLab, Orsay, France
[13α] INFN, Sezione di Padova, Padova, Italy
[13β] Dipartimento di Fisica e Astronomia, Università di Padova, Padova, Italy
[14] Institute of Particle and Nuclear Physics, Charles University, Prague, Czech Republic
[15] Department of Physics, Pontifícia Universidade Católica do Rio de Janeiro, Rio de Janeiro, Brazil
[16] RCNS, Tohoku University, Sendai, Japan
[17] Université de Strasbourg, CNRS, IPHC, Strasbourg, France
[18] Centro de Astropartículas y Física de Altas Energías (CAPA), Universidad de Zaragoza, Zaragoza, Spain
[a] Université de Paris Cité, CNRS, APC, Paris, France

———————————————

[*] Email: anatael@in2p3.fr
[†] Email: mchen@queensu.ca
[‡] Email: mantovani@fe.infn.it
[§] Email: andrea.serafini@pd.infn.it
[¶] Email: strati@fe.infn.it
[‖] Deceased

The formation of the Earth remains an epoch with mysterious puzzles extending to our still incomplete understanding of the planet's potential origin and bulk composition. Direct confirmation of the Earth's internal heat engine was accomplished by the successful observation of geoneutrinos originating from uranium (U) and thorium (Th) progenies, manifestations of the planet's natural radioactivity dominated by potassium ($^{40}$K) and the decay chains of uranium ($^{238}$U) and thorium ($^{232}$Th). This radiogenic energy output is critical to planetary dynamics and must be accurately measured for a complete understanding of the overall heat budget and thermal history of the Earth. Detecting geoneutrinos remains the only direct probe to do so and constitutes a challenging objective in modern neutrino physics. In particular, the intriguing potassium geoneutrinos have never been observed and thus far have been considered impractical to measure. We propose here a novel approach for potassium geoneutrino detection using the unique antimatter signature of antineutrinos to reduce the otherwise overwhelming backgrounds to observing this rarest signal. The proposed detection framework relies on the innovative *LiquidO* detection technique to enable positron ($e^+$) identification and antineutrino interactions with ideal isotope targets identified here for the first time. We also provide the complete experimental methodology to yield the first potassium geoneutrino discovery.

Despite the primary constituents of the Earth being well-established and constrained [1], our inability to access the deep interior of our planet makes it impossible to measure its exact bulk composition directly. The quantification of the chemical elements present requires reliance on compositional models, which are still under great unsettled debate. Among Earth's trace elements, there are the so-called *heat-producing elements*[1], uranium, thorium and potassium, long-lived radioactive elements whose decay has produced heat since the formation of our planet.

Knowing the abundance of the Earth's heat-producing elements is fundamental to understanding the extent natural radioactivity contributes to the Earth's internal *heat power* of (47 ± 2) TW [2], wherein the remaining portion is due primarily to the *secular cooling* of our planet. While abundances of *refractory lithophile elements* (U, Th) are well constrained by observations in *chondrites*, the silicate Earth seems strongly depleted in *volatile elements*, such as potassium (K). At the present time, potassium is thought to contribute to ~20% of the radiogenic heat produced inside our planet [3], and, moreover, it is believed to have played a crucial role in the early days of Earth's formation. Indeed, because of the relatively higher decay rate of $^{40}$K ($t_{1/2}$ = 1.25 Gyr), due to its shorter half-live relative to $^{238}$U and $^{232}$Th, its contribution to the overall radiogenic heat may have reached up to ~50% during the early stages of Earth's history.

The abundance of K in the silicate Earth spans a factor ~2 among compositional models, ranging from 130 to 280 µg/g [4]. Our planet, however, exhibits ~1/3 to ~1/8 of its predicted potassium content when compared to chondrites. The *missing K* relative to the prediction could be due to the loss of K into space during planetary accretion [1] or segregation into the differentiating Earth's *core* [5]. Solving the intriguing riddle of the missing K with a direct measurement is crucial and would be a breakthrough in the comprehension of Earth's origin and composition, providing key tests of bulk Earth compositional paradigms and models.

The missing K problem is linked with another open question in Earth Science: the missing Ar [6]. The present amount of $^{40}$Ar measured in the atmosphere is approximately half of that produced within the Earth since its formation [7]. More than 99% of terrestrial argon is produced by $^{40}$K decay; estimates of the bulk mass of K in the solid Earth thus determine the amount of $^{40}$Ar degassed into the atmosphere. Since ratios of volatiles (e.g. H/$^{40}$Ar) are constant for a wide variety of rocks, the constraint on the amount of argon drives the understanding of the behaviour of other volatile elements (i.e. H, N, C and noble gases) during planetary formation and evolution, including water [8]. A fuller understanding would be possible by resolving the mysteries behind the missing fractions of K and Ar.

---

[1] Hereafter, all terms in *italics* will be defined in the Glossary.

While powering the internal energetic processes in the Earth, beta minus decays of the heat-producing elements lead to antineutrinos and heat, in a well-fixed ratio. A direct measurement of this "geoneutrino" flux at the surface, ~$10^6$ cm$^{-2}$ s$^{-1}$ [9], provides an effective method for exploring Earth's inaccessible interior with unique and unrivalled insights [10, 11]. In 2005, the KamLAND experiment (Japan) reported the first detection of the *geoneutrinos* [12]. Soon after, the Borexino detector (Italy) made subsequent measurements [13, 14]. Future observations are expected from SNO+ (Canada) [15] and JUNO (China) [16]. In all these experiments, the detection mechanism is based on the *Inverse Beta Decay* reaction on free protons, denoted as IBD(p). The interaction, $\bar{\nu}_e + p \rightarrow n + e^+$, provides a delayed coincidence between the e$^+$ and the neutron signals, correlated by a short time interval (typically τ ≈ 200 μs), and grants a distinguishing event signature with strong background rejection power. However, this reaction has an energy threshold of 1.806 MeV; hence only the geoneutrinos originating from the $^{238}$U and $^{232}$Th decay chains are detectable via the IBD(p) reaction since $^{40}$K geoneutrinos have a maximum energy of 1.311 MeV (Figure 1).

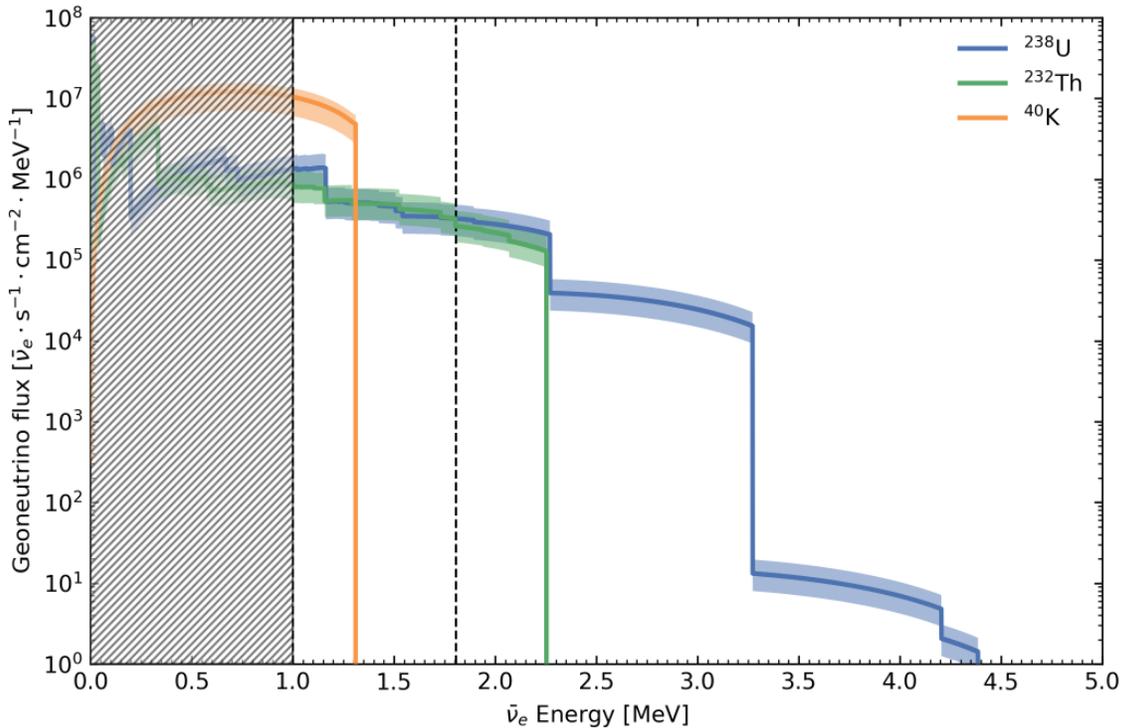

**Figure 1 | Geoneutrino Energy Spectra.** The $^{238}$U (blue), $^{232}$Th (green) and $^{40}$K (orange) geoneutrino fluxes expected at Laboratori Nazionali del Gran Sasso, as an example site, as a function of the antineutrino energy. The shaded lines show the variability range (see Methods section for details) due to the isotopes' masses and distributions in the Earth according to the models described in the Methods section, which corresponds to about ±50% uncertainty in the flux prediction. The black vertical dashed line at 1.806 MeV represents the energy threshold of the Inverse Beta Decay reaction on free protons, i.e. the reaction used by current liquid scintillator geoneutrino experiments. All contributions below the threshold are not measurable with today's technology. The region below 1.022 MeV, indicated by the black diagonal lines, represents the region which is not accessible with charged-current antineutrino capture on stable nuclear targets due to energy conservation in the reaction. The $^{238}$U spectrum above 3.272 MeV has traditionally not been shown because of its very low intensity.

The quest to detect $^{40}$K geoneutrinos requires another reaction to be identified. Antineutrino-electron scattering is a possible reaction with no energy threshold that can be considered for $^{40}$K geoneutrinos [17, 18]. The drawback, though, is that this interaction produces just a single recoiling electron signal that can be easily masked by irreducible backgrounds arising from both solar neutrinos, which are three orders of magnitude higher and also by natural radioactivity, such as beta minus decays or by Compton-scattered electrons produced by gamma rays. Suppression of those radioactivity backgrounds must far exceed the levels of ultra-low backgrounds achieved by Borexino [19] in order to have even a hope to detect $^{40}$K geoneutrinos. Further discussion on electron scattering interactions can be found in the Methods section.

Alternatively, instead of electrons as the target, one can consider charged-current weak interactions with atomic nuclei, as in the IBD(p) reaction, only with a target nucleus that is not hydrogen providing a reaction with a low enough energy threshold. This approach is also challenging since energy conservation establishes a minimum energy threshold of 1.022 MeV (twice the electron rest mass) for stable isotope targets, and thus the energy range of detectability for $^{40}$K geoneutrinos spans the limited range [1.022,1.311] MeV. The number of possible reactions meeting this condition is very limited. On the other hand, this type of interaction holds a unique advantage: geoneutrinos are antimatter. Indeed, the weak interaction leads to a one-to-one correlation between the production of antimatter (the geoneutrino) for each single β$^-$ decay of matter from Earth's heat-producing elements. The antimatter nature of geoneutrinos is evidenced by the manifestation of the positron (e$^+$) in the final state, as in the IBD(p) reaction; all charged-current weak interactions by electron antineutrinos produce positrons in the final state. Today's detectors cannot easily exploit this unique "antimatter signature" with its potentially resilient background rejection power. In today's technology, this scenario may only be possible in segmented detectors in an attempt to identify (tag) the annihilation gamma rays produced by the e$^+$, as was coarsely done in the first neutrino detection [20]. Despite recent efforts [21], this remains a pending challenge. Clearly, identifying the e$^+$ signature with high efficiency may provide the key to an experimental strategy for the first detection of $^{40}$K geoneutrinos.

Charged-current IBD antineutrino interactions $\bar{\nu}_e + {}^{A}_{Z}X \to e^+ + {}^{A}_{Z-1}Y$, hereafter referred to as IBD($^A$X), offer several possible target candidates for which the energy threshold is lower than 1.3 MeV, including those listed in [22], a study that focused on radiochemical-based detection. Cadmium, not listed in [22], was later proposed [23] as a promising new candidate isotope for $^{40}$K geoneutrino detection. IBD($^{106}$Cd) produces $^{106}$Ag and an e$^+$. The produced $^{106}$Ag subsequently decays, in turn, with an electron capture/β$^+$ decay branch in which an e$^+$ is emitted 59% of the time. Thus, $^{40}$K geoneutrino capture on $^{106}$Cd can generate *two* e$^+$ in a detector, delayed on average by the 24-minute $^{106}$Ag half-life. Despite the long coincidence time, this double-e$^+$ signal, correlated in space and time, offers a very distinctive signature for $^{40}$K geoneutrino detection that would strongly suppress backgrounds.

However, the limitation of cadmium as a potential target is the natural abundance of this interesting isotope candidate, $^{106}$Cd. At only 1.25%, the ability to build a detector with a huge number of $^{106}$Cd nuclei is unrealistic, given the anticipated size of the detector required to detect $^{40}$K geoneutrinos and the rather prohibitive cost of isotopic enrichment, with current technology, at the required scale. Consequently, attention must be turned to other possible targets with high natural abundance. For most of the possible candidates, the IBD($^A$X) reactions do not offer a delayed coincidence signature as in the case of IBD(p) and IBD($^{106}$Cd), and one must then ponder whether the clear identification of a single e$^+$ in a detector could be distinctive enough for robust detection of $^{40}$K geoneutrinos. Could the unique antimatter signature of geoneutrinos be reliably exploited for efficient detection and background control?

The scientific and technological development foreseen for detecting these low-energy antineutrinos, together with their relevance in Earth sciences, make the $^{40}$K geoneutrino measurement the "holy grail" for neutrino geoscientists. For exploring this low energy region of the geoneutrino spectrum (Figure 1), considered today as being practically impossible, a set of new reactions meeting all the necessary conditions, such as energy threshold (<1.3 MeV), natural abundance, and cross section, are discussed in this article, together with a possible experimental detection technique which features the unambiguous identification of single e$^+$ signals in a detector that greatly suppresses today's most limiting backgrounds.

The *LiquidO* detection technique [24] enables the clear identification of e$^+$ in a detector arising from both the spatial topology of the event and the time pattern for energy deposition and light collection. The unique double-e$^+$ signal of the $^{106}$Cd geoneutrino reaction combined with the ability of *LiquidO* to unambiguously detect e⁺ initially prompted this study of ⁴⁰K geoneutrinos using *LiquidO*. Our studies henceforth further focused on a new, full methodology exploiting the unique antimatter signature as the essential fact. We therefore examined several potential candidate interactions that lead to a single-e⁺ manifestation to enable practical detection of ⁴⁰K incident geoneutrinos for the first time. Geoneutrinos interact in a detector and produce antimatter (e⁺). The identification of an antimatter annihilation signal in a detector in and of itself strongly suppresses other backgrounds (such as from natural radioactivity) caused mostly by matter particles. Though there are some potential sources of background that can produce true e⁺ signals (or β⁺) in a detector, as discussed further in this article, they are much rarer than those that produce single electrons (or β⁻), as discussed before. *LiquidO*'s approach provides both a distinct e⁺ event signature as well as the means for deploying potential promising targets for ⁴⁰K geoneutrino detection via high-level detector doping, thanks to its opacity-based detection medium.

**Table 1 | The Most Promising Inverse Beta Decay (IBD) Target Isotopes and Expected Geoneutrino Signals.** The 1st column lists the target and product atoms involved in the IBD reaction, the 2nd column the IBD target Isotopic Abundance (IA) in percentage, the 3rd column the IBD reaction energy threshold ($E_{th}$) in MeV, the 4th column the Log(ft) value for the corresponding β decay of the final state, which is a measure of the reaction cross section. The IA, $E_{th}$ and Log(ft) values are taken from the ENSDF database [25], whose specific references are found in the 5th column. The 6th, 7th and 8th columns report, as central values, the expected geoneutrino signal in the [$E_{th}$ ; 3.272] MeV energy range respectively from uranium, thorium and potassium distributed in the Earth's lithosphere and mantle. The hypothetical location of the Laboratori Nazionali del Gran Sasso has been chosen as example site. The range in square brackets provides the variability of the expected predicted signal, thus defining the "minimal" and "maximal" scenarios for the masses and distributions of heat-producing elements in the Earth determined with different Bulk Silicate Earth compositional models. See Methods for further details. The expected geoneutrino signals are given in Terrestrial Neutrino Units (TNUs), corresponding to a 1-year acquisition time and $10^{32}$ atoms for each chemical species (i.e. a number of IBD target atoms corresponding to $10^{32}$ scaled by the isotopic abundance) [2]. Chlorine (³⁵Cl) and copper (⁶³Cu) represent the most promising targets for ⁴⁰K geoneutrinos detection, while hydrogen (¹H), and cadmium (¹⁰⁶Cd) are here reported for comparison. A full list of the suitable target isotopes for ⁴⁰K geoneutrinos detection can be found in the Methods section.

| Target process | IA [%] | $E_{th}$ [MeV] | Log(ft) | Ref | S(U) [TNU] | S(Th) [TNU] | S(K) [TNU] |
|---|---|---|---|---|---|---|---|
| ¹H → ¹n | 99.99 | 1.806 | 3.0170 | [26] | 31.5 [24.0 ; 47.0] | 9.0 [6.4 ; 14.1] | / |
| ⁶³Cu → ⁶³Ni | 69.15 | 1.089 | 6.7 | [25] | 0.85 [0.64 ; 1.26] | 0.49 [0.35 ; 0.77] | 0.10 [0.07 ; 0.13] |
| ⁶³Cu → ⁶³Ni* | | 1.176 | 5 | [22] | | | |
| ³⁵Cl → ³⁵S | 75.76 | 1.189 | 5.0088 | [27] | 0.73 [0.56 ; 1.09] | 0.43 [0.30 ; 0.67] | 0.10 [0.07 ; 0.13] |
| ¹⁰⁶Cd → ¹⁰⁶Ag | 1.25 | 1.212 | 4.1 | [28] | (1.7 [1.3 ; 2.6]) · 10⁻¹ | (9.7 [6.9 ; 15.2]) · 10⁻² | (5.1 [3.7 ; 6.6]) · 10⁻³ |

---

[2] The quantity $10^{32}$ atoms is, roughly speaking, the number of hydrogen atoms in 1 kiloton of CH₂-based liquid scintillator, with a native H fraction of 14.3% per unit of mass. This is the origin of the Terrestrial Neutrino Unit (TNU): equal to the number of IBD geoneutrino events per year with $10^{32}$ targets, which is used to quantify the detected geoneutrino signal.

# Candidate Targets for $^{40}$K Geoneutrino Detection

Table 1 lists the most promising target isotopes leading to a plausible IBD($^A$X) interaction with a low enough threshold to detect $^{40}$K geoneutrinos and high natural isotopic abundance. The estimated signal event rates are provided, including those of the IBD(p) and IBD($^{106}$Cd) for comparison. Since the crustal thickness and the amount of heat-producing elements present in the *crust* varies based on the location, the intensity of geoneutrino signals is strongly dependent on the detector site.

In the following calculations, the Laboratori Nazionali del Gran Sasso (Italy) was chosen as an example site, given the average crustal thickness (35 km) being an intermediate case between the extreme value of thinnest oceanic crust (e.g. Hawaii, 5 km) and the thickest continental crust (e.g. Himalayas, 70 km). Full details of the cross-section estimate, as inferred from the Log($ft$) value[3], rate calculations and a longer list of potentially suitable target isotopes for $^{40}$K geoneutrinos detection can be found in the Methods section, including arguments for discarding many of them.

Figure 2 shows the cross section and detected energy spectrum for several proposed targets, revealing that the most promising isotopes for detecting $^{40}$K geoneutrinos are $^{35}$Cl and $^{63}$Cu, after weighting the event rates by their isotopic abundance. The $^{40}$K geoneutrino signal rates are roughly the same for chlorine and copper, which are a factor of ~20 higher than the signal rate in the next best choice (cadmium), predominantly diminished by its poor natural abundance. Both Cl and Cu are further evaluated in our study as to their susceptibility to potential backgrounds and the ability to deploy either of them in a detector. The detection technique being proposed here is ideally suited for distinguishing e$^+$ signals and for doping with elements such as Cl and Cu, and will be described next before the specific discussion of Cl and Cu in the proposed *LiquidO* detector.

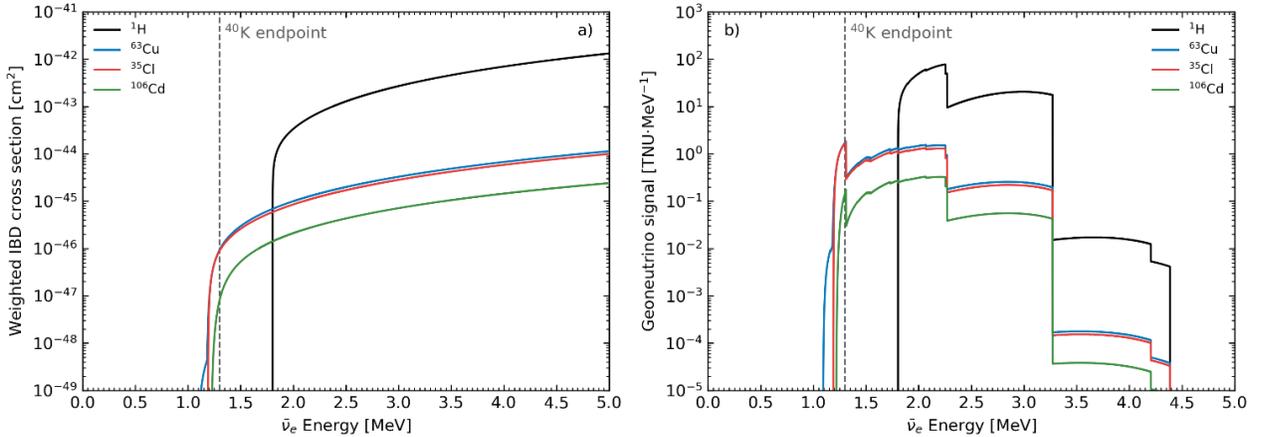

**Figure 2 | Cross Sections and Expected Geoneutrino Measured Spectra for the $^1$H, $^{35}$Cl, $^{63}$Cu and $^{106}$Cd Inverse Beta Decay (IBD) Targets.** Plot **a)** shows IBD reaction cross sections for the 4 different targets as a function of the incoming antineutrino energy, weighted by the corresponding target isotopic abundance (see Table 1). Plot **b)** shows the expected geoneutrino measured spectra at the Laboratori Nazionali del Gran Sasso (Italy), as an example site, originating from uranium, thorium and potassium distributed in the Earth's lithosphere and mantle, calculated as described in the Methods section. The expected geoneutrino spectra are given in Terrestrial Neutrino Units (TNUs), corresponding to a 1-year acquisition time and $10^{32}$ atoms for each chemical species. In both figures, the dashed vertical line indicates the endpoint of the potassium geoneutrino spectrum at 1.311 MeV.

---

[3] In nuclear beta decay, the half-life $t_{1/2}$ and the integral of the Fermi function $f$ quantify the degree of similarity between the initial and final nuclei involved in the transition. For inverse beta decay (IBD), the antineutrino absorption cross section involves the same nuclei (reversing initial and final states) and is inversely proportional to $ft_{1/2}$, often referred to simply as $ft$.

## LiquidO: a New Detection Technique

The *LiquidO* detection technique [24] has two features making it ideal for the detection of $^{40}$K geoneutrinos. The first is that the requirement of very high transparency for the liquid scintillator is relaxed in *LiquidO* because scintillation light gets collected using an array of closely spaced wavelength-shifting fibres distributed throughout the detector. Organic liquid scintillators can be doped with various elements [29]; traditionally, this has potentially affected the transparency and intrinsic light yield of the resultant liquid scintillator cocktail. *LiquidO*'s tolerance to reduced transparency opens up the possibility to dope detectors with different elements to higher loading fractions. *LiquidO* is, therefore, well aligned with the main objective of identifying the best isotope candidates to be used as targets for charged-current reactions with low-energy $^{40}$K geoneutrinos. As a by-product, the same technique can be used for the detection of lower-energy antineutrinos originating from reactors, or nuclear fuel, or waste sites. In turn, reactors may be used as effective "test-beam" facilities to experimentally demonstrate much of the described methodology, thus benefiting from its high and exponentially increasing antineutrino flux at lower energies.

The second feature of *LiquidO* applicable to geoneutrino detection is its powerful particle identification capability. This comes from both the topology of the energy deposited in the events and from the time pattern for the light, proportional to the energy deposited, to be collected. This is referred to as *energy flow* in *LiquidO* (Figs. 3b, 3c). Positron-event topology can be imaged in *LiquidO* (Figure 3a), in which the e$^+$ central energy deposition is linked to its kinetic energy, followed by its two annihilation gamma rays that can be readily identified. Indeed, *LiquidO*'s ability to unravel the positron annihilation pattern is so striking that the technique is currently being explored for high performance imaging [30]. This capability makes *LiquidO* well-suited to detect antineutrinos, and their distinctive e$^+$ signature tag can be exploited for IBD($^{35}$Cl) and IBD($^{63}$Cu) observation, perhaps even without the need for a delayed coincidence.

## Methodology

It is instructive to first consider the detection of $^{40}$K geoneutrinos using a *LiquidO*-style detector, using IBD($^{35}$Cl). The copper case will be addressed after. Chlorine has the advantage that it can be easily loaded in an organic liquid scintillator, and there are several possible ways to achieve high doping, up to a hypothetical 50% in weight. For instance, a chlorinated-benzene compound could be used as a liquid scintillator in *LiquidO*. Several compounds, such as (but not limited to) dichlorobenzene ($C_6H_4Cl_2$), fluoresce and can function as (or together with) a liquid scintillator. Moreover, chlorine could also be doped in a typical liquid scintillator by mixing in a non-fluorescing chlorinated solvent such as tetrachloroethylene ($C_2Cl_4$). These and other different doping options would require dedicated R&D to fully develop feasible deployment scenarios.

Table 1 shows that the rate of IBD(p) events from U and Th geoneutrinos is much larger than IBD($^{35}$Cl) from $^{40}$K geoneutrinos. A chlorine-loaded detector that has abundant hydrogen, as in most liquid scintillators, will record many IBD(p) reactions and this serves to measure U and Th events with high statistics. This way, one measures the U and Th geoneutrino fluxes using the IBD(p) events, tagged by their distinctive neutron capture, to allow their contribution to the IBD($^{35}$Cl) events to be extracted and inferred, as illustrated in Figure 4.

In evaluating the concept of single e$^+$ identification in *LiquidO* with chlorine, to select IBD($^{35}$Cl) events, the central question is this: without a delayed coincidence, would a single e$^+$ be a robust enough signal compared to backgrounds? Even if e$^+$ events can be easily distinguished from normal $\alpha$, $\beta$ and $\gamma$ radioactive backgrounds, are there any backgrounds that produce actual e$^+$ events in a detector, and how do these background rates compare to the anticipated $^{40}$K geoneutrino signal's very low rate?

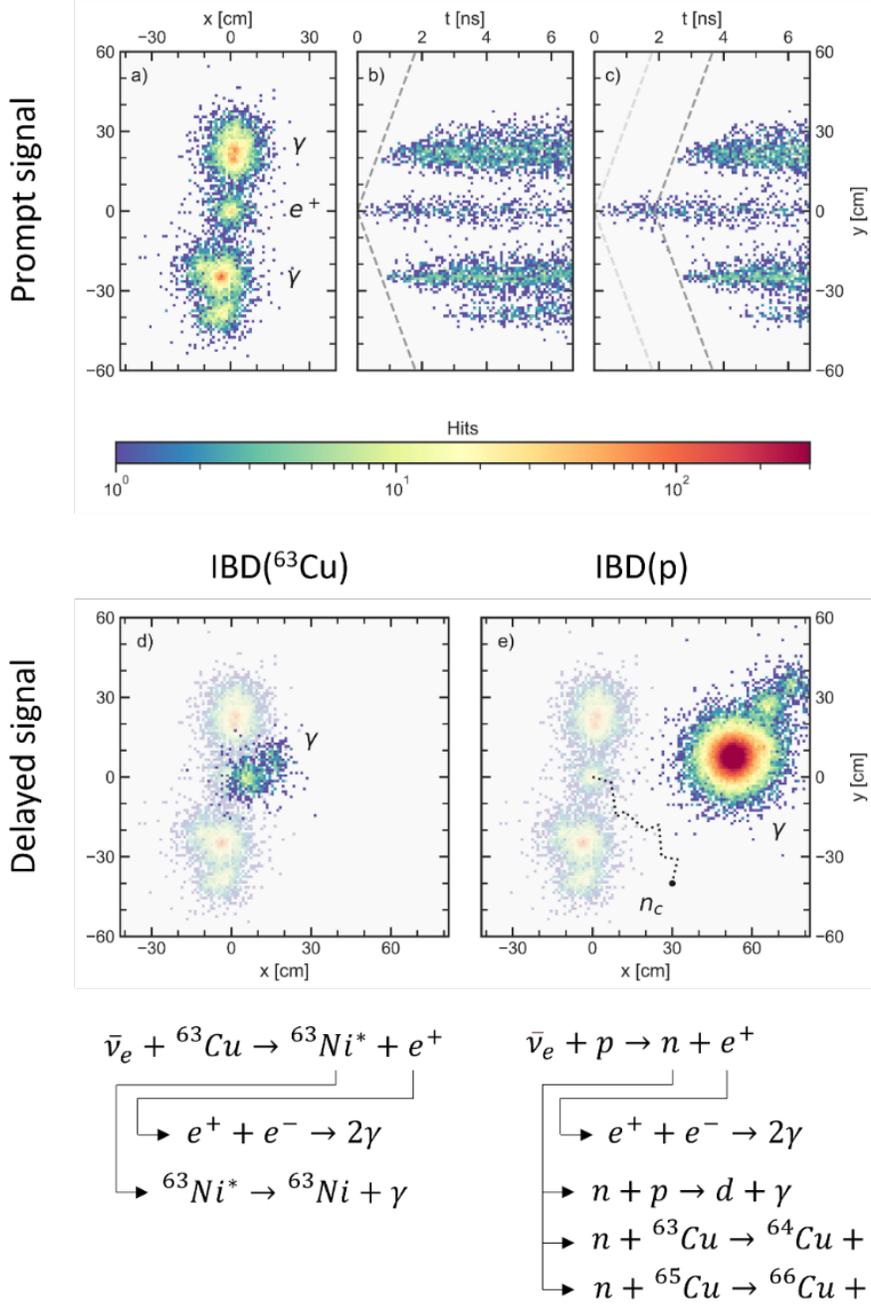

**Figure 3 | Simulated IBD Events in a LiquidO Detector Doped with Copper.** Plot **a)** shows the event topology of a 100-keV single $e^+$ featured by its emission of two back-to-back 511-keV annihilation gammas to be used for event-wise tagging. This topology is common to IBD(p) and IBD($^{63}$Cu) events upon the $e^+$ emission. Plot **b)** shows the time pattern for the scintillation light (energy) to be collected in LiquidO for the $e^+$ event shown in a), assuming that the annihilation of the $e^+$ is immediate, as opposed to the formation of ortho-positronium, depicted in plot **c)**, that typically delays the annihilation by a few ns. The topology and the time pattern of a positron annihilation are the key signatures for any IBD interactions, and they will be key tagging features for IBD($^{63}$Cu). The energy flow (i.e. the topology of the energy deposition as a function of time) reveals the two annihilation gamma rays propagating away from the point of the $e^+$ energy deposition. The scintillation light produced by subsequent multiple gamma ray Compton scattering diffuses with a speed lower than the speed of light (represented by the grey dashed lines) and is collected by nearby fibres and sums to 511 keV of energy deposited by each gamma ray. Plot **d)** shows the event topology of the delayed 87-keV gamma ray from the de-excitation of $^{63}$Ni* following a prompt $e^+$ signal of an IBD($^{63}$Cu) reaction. Last, plot **e)** shows the event topology of the delayed gamma ray produced in the neutron capture on hydrogen following a prompt $e^+$ signal of an IBD(p) reaction. The neutron that emerges from the point of interaction in d) captured at the point labelled $n_c$. Neutron captures on copper would also be easily identified in LiquidO because numerous gamma rays are emitted with a large total energy, with reaction Q-values of 7.9 MeV and 7.1 MeV for $^{63}$Cu and $^{65}$Cu, respectively. The presence of Cu also shortens the time coincidence between prompt ($e^+$) and the delayed neutron capture, allowing better accidental background reduction. The $e^+$ annihilation is shown faded in c) and d), representing the fact that it occurred prior to the depicted delayed signals. The LiquidO vertex position precision is expected to be at the sub-cm level, up to ~1 mm depending on the detector readout configuration, further improving accidental coincidence and cosmogenic background rejection via the correlation with a preceding cosmic muon track. All the plots have a common colour-coded z-scale indicating the raw number of photons hitting the LiquidO fibres, labelled "hits". The photon detection efficiency has not been included as it is specific to the detector configuration and optimisation studies are in progress.

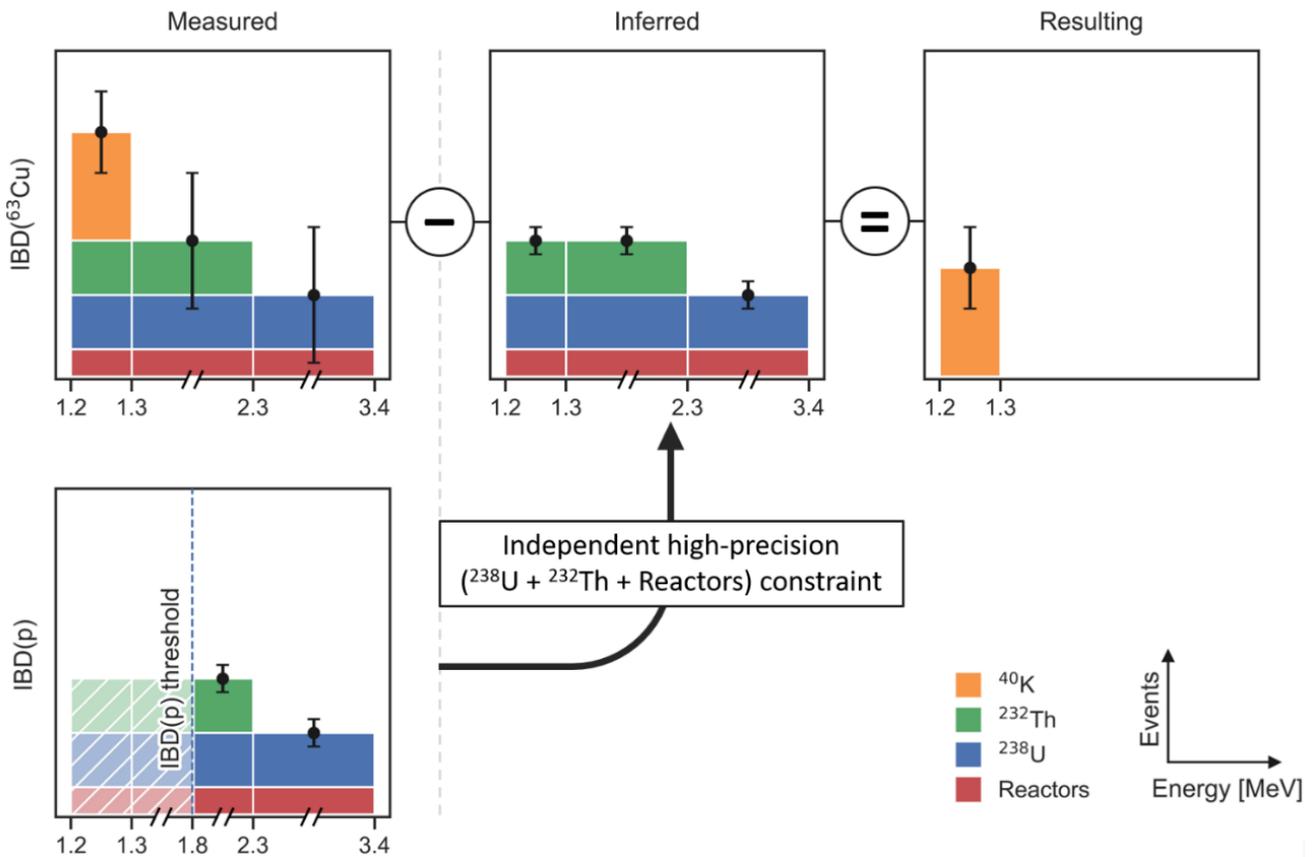

**Figure 4 | Methodology for $^{40}$K Geoneutrino Signal Extraction.** Reactor antineutrinos as well as $^{238}$U and $^{232}$Th geoneutrino fluxes are determined by high statistics measurements of IBD(p) events from which their contribution to IBD($^{63}$Cu) events can be inferred at the percent-level. Those same antineutrino contributions need to be subtracted over the [1.189;1.311] MeV geoneutrino energy range, as they are effectively irreducible background to the IBD($^{63}$Cu) based $^{40}$K geoneutrino signal. In this figure, the amplitudes and error bars of the histograms are not to scale and have been drawn for explanatory purposes.

## Backgrounds

There are indeed potential backgrounds that can produce genuine e$^+$ in the detector. They come from other antineutrinos (from nuclear power reactors and U/Th geoneutrinos), and from β$^+$ emitting background sources. The latter are considerably rarer than β$^-$ backgrounds, which helps make the problem somewhat tractable. Finally, there are gamma-ray interactions that can produce or mimic the e$^+$ signature. Table 2 summarises the nature of all backgrounds and how they might be dealt with and suppressed in a *LiquidO* detector.

All the antineutrino backgrounds, though irreducible, are easily managed. Using IBD(p) reactions in the liquid scintillator, both the reactor antineutrinos as well as the U and Th geoneutrinos can be well measured and constrained in the same detector with minimal or no reliance on predictions. Within the narrow energy range, above the IBD($^A$X) reaction threshold and below the endpoint (1.3 MeV), $^{40}$K geoneutrinos substantially dominate, and the contributions even from a relatively larger reactor antineutrino flux will be insignificant (excluding sites that are very close to nuclear reactors).

**Table 2 | Backgrounds to the $^{40}$K Geoneutrino e$^+$ Signal in LiquidO.** The $^{40}$K geoneutrino signal is a single e$^+$ that can be identified in LiquidO, as illustrated and described in Figure 3. All possible backgrounds, including irreducible antineutrinos, true e$^+$ sources and gamma rays faking e$^+$ events are listed in the table, along with a brief note how these backgrounds compare and how they might be further suppressed. Further details can be found in the main text.

| e$^+$ backgrounds | | |
|---|---|---|
| Source | Expectation | Suppression |
| IBD($^A$X) of U and Th geoneutrinos | ~1/5 rate of predicted $^{40}$K geoneutrinos in K energy interval | Irreducible but measured by IBD(p) |
| IBD($^A$X) of reactor antineutrinos | ~1/60 rate of predicted $^{40}$K geoneutrinos in K energy interval | Irreducible but measured by IBD(p) |
| β$^+$ decays of naturally occurring $^{40}$K in the detector | ~10$^{-5}$ e$^+$ per year per ton | Use of decay energy spectrum to constrain |
| β$^+$ decays of cosmogenic isotopes | Similar to β$^+$ isotope production in current experiments | Cosmic muon veto and energy spectrum for further suppression |
| Pair production (e$^+$-e$^-$) by gamma rays (conversion) | Pair production cross section in the [1.022; 1.144] MeV range is ~10$^{-3}$ times the Compton scattering cross section that dominates in current experiments | Gamma rays make other interactions prior to converting; *LiquidO* event pattern helps reject |
| Multiple Compton scattering (fake e$^+$ signal) | e$^+$ signal is still very distinctive compared to gamma-ray-induced single electron recoils | Monte Carlo study of rejection of gamma rays using event pattern and energy flow |

Figure 4 illustrates the method for the quantification and statistical subtraction of this irreducible background. The excess of e$^+$ events in the proper energy range, above the predicted background from U, Th geoneutrinos and reactor antineutrinos, is expected to be a clean sample of $^{40}$K geoneutrinos.

A minor additional contribution comes from the neutrinos and antineutrinos produced in pionic and muonic decays in the atmosphere, typically referred to as atmospheric neutrinos, which can also interact in many ways with the nuclei constituting the scintillator and the doping isotopes. Given the signal of a low-energy single positron, the main contribution from atmospheric neutrinos is expected to be led by electron-flavoured antineutrinos undergoing the IBD($^A$X) reactions. In this context, only neutrinos whose energy is very close to the reaction energy threshold would be important. Interactions on C have also been considered. Given the tight signal topology, narrow energy acceptance constraints, and the fact that the atmospheric neutrino flux falls steeply with energy [31], this background is determined to be negligible. The estimates from running experiments without topological e$^+$ tagging, such as Borexino and KamLAND [32], confirm a negligible impact.

Of greater concern are β$^+$ decaying backgrounds. In natural radioactivity, there is perhaps only one important β$^+$

emitter, that being $^{40}$K itself. Low background physics experiments have learned how to eliminate this background to achieve very low levels, with the primary concern typically being the 1.46 MeV gamma ray that is emitted following the electron capture decay branch 10.67% of the time. In comparison, the β$^+$ decay branching ratio is 10$^{-5}$. With the remarkably low levels of $^{40}$K achievable in a liquid scintillator [19], one can estimate this single e$^+$ background to be less than one such β$^+$ decay per kiloton of detector per 100 years. The energy spectrum of the e$^+$ background from $^{40}$K in the detector extends to slightly higher energies than the e$^+$ signals produced by $^{40}$K geoneutrino charged-current interactions on $^{35}$Cl or $^{63}$Cu. This provides another handle for discriminating this background, as less than 15% of the energy spectrum of the e$^+$ produced by $^{40}$K-β$^+$ decays in the detector overlaps with the $^{40}$K geoneutrino spectrum. The concentration of $^{40}$K in the scintillator can also be precisely estimated through the observation of the much more common 1.46 MeV gamma-ray emission and then used to derive its abundance *in-situ*, in order to subtract its expected e$^+$ background in the $^{40}$K geoneutrino energy region.

An additional possible source of β$^+$ emitters is represented by cosmogenic isotopes produced by cosmic-ray interactions with materials in the detector [33], especially while those materials are on the surface. Short-lived isotopes that are produced can, in principle, be vetoed by identifying the cosmic-ray muon passing through the detector and applying a position-time veto window in which subsequent events are rejected. Tracking muons with high precision will be possible with *LiquidO* and is expected to be very effective. However, long-lived isotopes are also a possibility and, if an excessive amount were produced at the surface, it would require an extreme campaign, almost necessarily to be conducted underground, to purify materials to remove the activated cosmogenic isotopes, if that's at all possible. As we will examine later, cosmogenic β$^+$ emitters can represent a major concern for the candidates identified for the detection of $^{40}$K geoneutrinos.

To complete the discussion of e$^+$ backgrounds one must consider gamma rays and how they could mimic the sought-after e$^+$ signature. Gamma rays can produce true e$^+$ in a detector through pair production. The subsequent annihilation of the e$^+$ produces the expected event topology in *LiquidO*. The $^{40}$K geoneutrinos produce e$^+$ with very little kinetic energy, between [0, 0.289] MeV. The pair production cross section on Cu and Cl is ~10$^{-3}$ times that of Compton scattering at threshold plus a few hundred keV [34]. Since pair production is suppressed, it seems feasible that energy information of the candidate e$^+$ event together with requiring no previous interaction of the gamma ray prior to it undergoing pair production could be sufficient to further reject this background. The specific quantification of this background requires detailed simulation-based optimisation for different doping scenarios that consider the radiation properties of the overall medium. A detailed analysis of this nature will be addressed in future feasibility studies.

The final background that was considered is from multiple Compton scattering of gamma rays that could, by chance, mimic the e$^+$ signature. The energy flow information that is accessible in an event in *LiquidO* would strongly suppress this background. A true e$^+$ event originates the two 511 keV annihilation gamma rays from the same point, propagating with the speed of light. Multiple gamma-ray Compton scatters would tend not to produce these two simultaneous and then displaced energy depositions with the topology and resolution granted by *LiquidO*, as illustrated in Figure 3. Furthermore, the antineutrino-produced positron can also lead to the non-negligible formation of positronium, i.e. an unstable exotic atom with a loosely-bound electron [35]. Both the formation and stability of positronium depend on the type of medium [36], including the possible presence of doping [37]. This leads to possible variations in the formation rate and annihilation pattern exhibited mainly due to the production of ortho-positronium [35] with differences in lifetime and number (e.g. three) of annihilation gammas. In traditional scintillators, the ortho-positronium formation fraction can be as high as ~50%, with a decay time of order ≤3 ns [36], consistent with previous, precise observations in antineutrino experiments [38]. The rich positronium pattern may provide an extra handle for background control in a *LiquidO*-based detector relying heavily on positron tagging (Figure 3). The expected positronium pattern can be experimentally characterized with extreme precision in the laboratory for any given scintillator configuration [36, 37].

We conclude this discussion of e$^+$ backgrounds by postulating that a future experiment might be able to control gamma-ray backgrounds sufficiently well, with the antimatter signature of e$^+$ annihilation then being sufficiently distinguishing. The feasibility of this approach depends heavily on *LiquidO*'s detection and doping performance, still under experimental validation. The potential detection of $^{40}$K geoneutrinos appears *a priori* feasible.

This brings us to finally discuss possible backgrounds specific to the two elements identified for detecting $^{40}$K geoneutrinos: copper and chlorine. For the case of chlorine as a geoneutrino target, despite having one of the highest cross sections among the possible targets and being simpler to dope in a liquid scintillator to a high weight fraction, there is a problematic cosmogenic background. Natural chlorine contains trace amounts of the radioisotope $^{36}$Cl, at the part in 10$^{13}$ level. This isotope is produced in the atmosphere by spallation, upon cosmic-ray interactions on $^{36}$Ar, and in the upper part of the *lithosphere* by thermal neutron activation of $^{35}$Cl. With a half-life of 3×10$^5$ years, and a β$^+$ decay branch with 0.02% probability, this isotope would produce ~10$^{10}$ positrons per year per 10$^{32}$ chlorine atoms. This would, unfortunately, completely overwhelm the $^{40}$K geoneutrino expected event rate (~0.1 TNU). Chemical purification does not separate chlorine isotopes, and large-scale isotopic separation is not feasible. For this reason, chlorine loaded in a *LiquidO* detector to search for $^{40}$K geoneutrino interactions producing a single e$^+$ signal has to be ruled out. The authors have highlighted this otherwise promising candidate to note the importance of carefully considering cosmogenic production of trace isotopic backgrounds in rare event searches. The low event rate and possible impact of unknown combinatory backgrounds strongly favour identifying another candidate target element with a more robust signal topology, if possible, while not suffering from having a long-lived cosmogenic β$^+$ emitter.

## Copper – The Ideal Candidate and the Only One?

We next discuss copper as a candidate for doping in *LiquidO* for $^{40}$K geoneutrino detection, as it has an expected event rate as high as chlorine. Doping copper in an organic liquid scintillator has not been previously accomplished. Nevertheless, techniques exist for producing scintillator cocktails that are miscible with water. Copper in an aqueous solution could be loaded in this manner. Organocopper compounds also exist, and the chemistry of copper is arguably more varied than that of chlorine. The *LiquidO* approach that employs an opaque scintillator, with a short light-scattering length, is amenable to new metal-loading approaches (e.g., dispersion of nanoparticles). At 69% natural isotopic abundance, $^{63}$Cu would be the specific isotope that has a low enough energy threshold for IBD($^{63}$Cu) to be usable to detect $^{40}$K geoneutrinos.

The ground state of $^{63}$Cu has spin-parity 3/2$^-$. IBD($^{63}$Cu) reactions would lead to the $^{63}$Ni ground state (1/2$^-$) or $^{63}$Ni$^*$ excited state (5/2$^-$) at the 87 keV energy level; both would be allowed transitions such that their cross sections are expected to be favourable. The cross section for the transition to the $^{63}$Ni ground state can be estimated from $^{63}$Ni β decay, using its Log(*ft*) value from the half-life, branching ratio and decay spectrum. On the other hand, for transitions to the $^{63}$Ni$^*$ excited state, there are no previous measurements of the β decay partial half-life for this nuclear state. We have assumed a typical Log(*ft*) value of 5, as was done in [22], noting that the Log(*ft*) value to the $^{63}$Ni ground state transition is worse (Log(*ft*) = 6.7). The $^{63}$Ni$^*$ excited state does have a higher Q-value and thus its associated β decay half-life should be shorter, leading to a smaller Log(*ft*) value and to a higher cross section. One would need to confirm or measure this cross section estimate for IBD($^{63}$Cu) to the excited state using a new measurement of $^{63}$Ni$^*$ β decay – maybe exceedingly rare – or perhaps with arguments from nuclear theory, possibly in conjunction with experimental results from $^{63}$Cu(n,p) nuclear reactions.

If IBD($^{63}$Cu) transitions to $^{63}$Ni$^*$ have a favourable enough cross section to make it a candidate, it has another advantage. The excited state has a lifetime of 1.67 μs for decaying by emission of a gamma ray with 87 keV energy. This would provide a delayed coincidence that is exploitable in rejecting all backgrounds, including true e$^+$ backgrounds. Explicitly, IBD($^{63}$Cu → $^{63}$Ni$^*$) would first produce a clear e$^+$ signal in a *LiquidO* detector, followed by the detection of an 87 keV gamma ray correlated in time and space with the e$^+$. Figure 3d illustrates the delayed detection of an 87 keV gamma ray in *LiquidO*. Given the desired high loading fraction of copper, photoelectric absorption (dependent on atomic number) of the 87 keV gamma ray would be most probable, and the expected signal would be a single-point energy deposition of 87 keV. The specific energy and the isolated event topology, near the IBD(Cu) e$^+$, would make this delayed event extremely distinctive, leading to an overall major improvement in the robustness of the detection.

Cosmogenic backgrounds associated with copper may have some advantages relative to chlorine. Copper does not have long-lived β$^+$ decaying isotopes. There is $^{64}$Cu that can be produced by neutron activation, which can decay with both β$^-$ and β$^+$ emission, with a branching ratio of 17.6% for β$^+$ decay. Thus, the control of neutrons would be important and might imply overburden constraints for the detector that are not expected to be onerous.

Potential sources of neutrons in a detector are: (i) cosmic-ray induced neutrons produced in the atmosphere impinging on the detector, (ii) muon-induced neutrons generated in electromagnetic/hadronic interactions of fast muons or by muon capture in nuclei contained in or surrounding the detector (iii) neutrons created in (α,n) reactions and those following the spontaneous fission of the U in the rocks around the detector's experimental hall and (iv) neutrons produced as a result of the IBD(p) process itself. Cosmic ray-induced neutrons do not represent a major concern as they can be shielded by placing the detector below the Earth's surface. In an underground neutrino detector, their flux would be practically zero, with the only concern represented by the exposure of the liquid scintillator to the surface neutron flux before detector filling. As the half-life of $^{64}$Cu is 12.7 hours, after exposing copper to a non-negligible neutron flux (e.g., the surface cosmic ray neutron flux), it would only be required for the copper to "cool" for a period of several days (~11 days would provide a factor >10$^6$ reduction). The half-life is short enough for this to be an effective strategy, and after decaying away, the copper-loaded detector would just need to be surrounded by thermal neutron shielding to minimize continued production of $^{64}$Cu. Passive or active shielding (e.g., a water pool) would also prove effective in reducing the flux of neutrons produced by muons interacting with the rocks surrounding the detector. Cosmic-ray muons and muon-induced neutrons may produce some $^{64}$Cu nuclei; however, muons are easily and efficiently tagged in neutrino detectors, especially in *LiquidO*, where active vetoing techniques can be employed to further suppress $^{64}$Cu-backgrounds via temporal and spatial fiducial cuts around a passing muon track. Even for a detector that is shielded and underground, cosmic-ray muons might not be completely suppressed and would generate some equilibrium level of $^{64}$Cu in a detector. A future experiment would need to be designed to minimize and/or efficiently tag muon-induced neutrons that might activate $^{64}$Cu at some residual level.

The last source of neutrons to consider originates from the IBD(p) reaction itself. When interacting with an antineutrino, proton targets contained in the scintillator produce a neutron in the final state. This neutron is critical for the identification of IBD(p) reactions. Any misidentification could lead to a confusion between the IBD(p) and IBD($^{63}$Cu) detection channels, making it effectively impossible to detect IBD($^{63}$Cu) events, which are statistically disfavoured by several orders of magnitude, as shown in Table 3. Both naturally occurring copper isotopes, $^{63}$Cu and $^{65}$Cu, have neutron capture probabilities much higher than hydrogen [39, 40] with $^{63}$Cu having the highest cross section [41]. Neutron captures on copper are easily identified in *LiquidO* because of the numerous gamma rays emitted in the process, with a large total energy (i.e. reaction Q-value) of 7.9 MeV and 7.1 MeV for $^{63}$Cu and $^{65}$Cu respectively [41]. In this case, the produced $^{64}$Cu β$^+$ emitters can be spatially and temporally vetoed on the basis of the detected gamma rays following the $^{63}$Cu neutron capture. To recapitulate, if an IBD(p) event produces a neutron that is captured, for example, on $^{63}$Cu, the IBD(p) is not missed because of the distinctive neutron-capture gamma signal. If the produced $^{64}$Cu later undergoes β$^+$ decay, this can be associated with the previous IBD(p) event's neutron-capture position and thus this e$^+$ is not counted as a background event for the IBD(Cu) signal.

**Table 3 | Expected Number of Geoneutrino and Reactor Antineutrino Events.** Antineutrino events estimated using a LiquidO detector located at the Laboratori Nazionali del Gran Sasso (Italy), as an example, and having a mass of 240 kton with 50% Cu loading and considering 10 years of data taking. Full details regarding the calculation of the geoneutrino and reactor antineutrino events rates are described in the Methods section.

| Detection reaction | Energy range | Events / 10 yr / 240 ktons | | | |
| --- | --- | --- | --- | --- | --- |
| | | Reactors | $^{238}$U | $^{232}$Th | $^{40}$K |
| IBD(p) | [1.806 – 3.27] MeV | 19530 | 27750 | 7948 | / |
| IBD($^{63}$Cu→$^{63}$Ni$^*$) | [1.176 – 1.311] MeV | 0.2 | 1.1 | 1.1 | 11.7 |

## Detection Significance Estimation

To estimate the $^{40}$K geoneutrino detection significance in a future experiment, one can consider the Poisson probability distribution of the hypothetical observation of N or greater events compared to the null hypothesis that the experimental observation came from only backgrounds. This can be calculated as:

$$\sum_{n=N}^{\infty} P(n;\ \mu + \Delta\mu) \qquad (1)$$

where N is the number of observed events, μ is the mean expected number of background events in the null hypothesis, and the uncertainty of the total background Δμ (both statistical and systematic contributions) can be accounted for by adding it to the mean [42]. A more formal calculation using a likelihood approach yields similar results because of the small rates and uncertainties involved. Generally speaking, if the expected number of background events is uncertain but can be constrained by a separate measurement, then the background as a nuisance parameter in the likelihood function for a Poisson counting experiment can also be well constrained. It essentially reduces the significance calculation to counting statistics of the anticipated signal versus the possible fluctuations of the known background.

We consider the case of copper with identified delayed coincidences with the 87 keV gamma ray. Backgrounds such as pair production or fake e$^+$ signals from multiple Compton scattering are eliminated by the time-position-energy coincidence requirements. As argued above, we will assume that natural and cosmogenic radioactivity of β$^+$ emitters are also suppressed, measured and constrained, and ultimately also eliminated by the coincidence requirement.

Only the irreducible antineutrino backgrounds remain to be quantified, in comparison to the $^{40}$K geoneutrino signal. In this case, the separate measurement of the U and Th geoneutrino background in the $^{40}$K geoneutrino energy range comes from the IBD(p) events, as described previously. Due to the relatively high rate of the IBD(p) events in a hydrogen-rich liquid scintillator, the fractional uncertainty of this background contribution will be small in comparison. One of the keys here that yields maximal discovery sensitivity is the data-driven subtraction of the U+Th signal, whose accuracy cannot be achieved via predictions based on geological arguments alone.

The statistical and systematic uncertainties in the reactor background contribution in the $^{40}$K geoneutrino energy range must also be included. Sites with minimal reactor contributions are nonetheless favoured; but even if the flux is relatively high and even if those uncertainties are fractionally much larger, they will still have minimal impact since the absolute event rate from reactor events in the proper energy range is so much smaller, as shown in Table 3.

All systematic uncertainties (U, Th, and reactors) can be included in the Δµ term, in the calculation of the background expectation and Poissonian probabilities. The fact that U+Th geoneutrino backgrounds to the IBD($^{63}$Cu) events are directly measured by IBD(p) events in the same detector helps as many systematics related to their relative rates cancel, enhancing the robustness of the proposed experimental methodology. There are some systematics that need to be included that affect the low-energy (≤1.8 MeV) IBD($^{63}$Cu) signal differently than the higher-energy IBD(p) signal, such as the reaction cross section uncertainties, U, Th and reactor spectrum shape uncertainties at those energies, and detector-related systematics that are energy dependent. Most of those systematics effects could be experimentally explored and validated by taking ancillary data close to a reactor with a small detector with the pertinent experimental setup, which would provide a large and known antineutrino flux for an experimental measurement of the $^{63}$Cu antineutrino capture cross section, especially the cross section to the excited state $^{63}$Ni$^*$. Ultimately, systematic uncertainties in all of the background estimations wind up having little impact on the $^{40}$K geoneutrino flux measurement (and its significance) since the $^{40}$K geoneutrino signal is nearly an order of magnitude larger than each of the U and Th geoneutrino backgrounds, as quantified in Table 3. The resulting $^{40}$K geoneutrino detection significance as a function of the detector mass is plotted in Figure 5, together with the corresponding uncertainty in the detection of U and Th geoneutrinos.

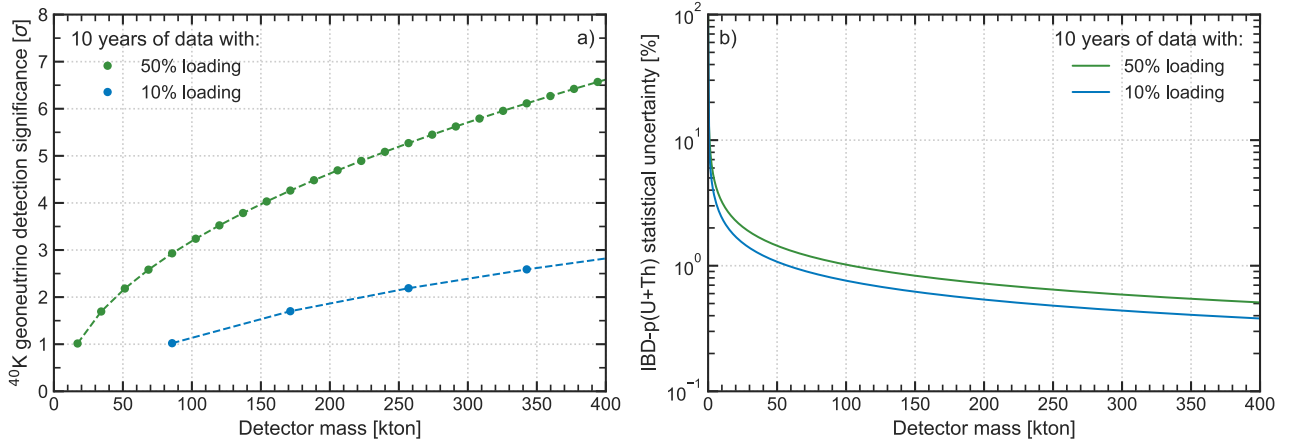

**Figure 5 | $^{40}$K Geoneutrino Detection Significance and Statistical Uncertainty on IBD(p) (U+Th) Signal.** Plots **a)** and **b)** refer to a Cu-loaded detector running for 10 years at the Laboratori Nazionali del Gran Sasso, as an example. The expected geoneutrino signal and the irreducible reactor and geoneutrino backgrounds in the IBD($^{63}$Cu) potassium energy region [1.176 ; 1.311] MeV are $S_{IBD(Cu)}$(K) = 0.10 TNU, $S_{IBD(Cu)}$(U+Th)= 1.9·10$^{-2}$ TNU, $S_{IBD(Cu)}$ (reactors)= 1.9·10$^{-3}$ TNU, as summarised in Table 3. The expected geoneutrino signal and the irreducible reactor and geoneutrino background in the IBD(p) energy region [1.806 ; 3.272] MeV are $S_{IBD(p)}$(U+Th)= 40.6 TNU[4], $S_{IBD(p)}$(reactors)= 22.2 TNU [43]. Two hypothetical Cu mass loadings are considered, 50% and 10% (green and blue lines, respectively), defined as the ratio between the Cu mass and the overall detector mass. Plot **a)** shows the $^{40}$K geoneutrino detection significance in number of σ and is calculated by conservatively considering a +1σ statistical uncertainty on the number of IBD(p) events, which is propagated in the estimation of the background events in the $^{63}$Cu potassium energy region (µ+Δµ), as defined in Equation 1. Each data point corresponds to an integer number of events N observed in the $^{63}$Cu potassium energy region. Plot **b)** shows the statistical uncertainty on the IBD(p) (U+Th) geoneutrino signal, which is estimated as the square root of the number of geoneutrino plus reactor IBD(p) events.

---

[4] This estimate is ~6 TNU higher with respect to [14], which adopts U and Th abundances inferred from local samplings [44]. Since this refined model does not report any information about K in the sedimentary deposits of the Apennines chain, here we calculate the expected geoneutrino signal using the global model of [45]).

The methodology for extracting the $^{40}$K geoneutrino signal (Figure 4) requires the estimation of the U and Th geoneutrino components plus the reactor backgrounds in the $^{40}$K geoneutrino energy window. Therefore, although *LiquidO* has single e$^+$ identification capabilities, the presence of two classes of events in such a detector, IBD($^{63}$Cu) and IBD(p), makes efficient neutron tagging a strict requirement. Any IBD(p) event that fails to reconstruct the accompanying neutron could be misidentified as an IBD($^{63}$Cu) event (or as a background e$^+$ event) that is missing the delayed 87 keV gamma ray from true IBD($^{63}$Cu→$^{63}$Ni$^*$), with the potential that those at low kinetic energies are mistaken for $^{40}$K geoneutrinos. *LiquidO* with copper is expected to have an enhanced neutron tagging efficiency since thermal neutron capture on copper is characteristically detectable, as described above. To summarise, IBD(p) events produce an e$^+$ signal and a neutron. The neutron must be efficiently identified so that IBD(p) events accurately measure the U and Th geoneutrino fluxes, as well as the reactor antineutrino background. Those neutrons are tagged by either the 2.2 MeV gamma ray following neutron capture by protons or by the energetic multiple gamma rays released following neutron capture by either of the stable isotopes of copper, $^{63}$Cu or $^{65}$Cu. If the neutron is missed, for example by undergoing (n,p) or (n,α) on Cu, with the final state particles not detected, it must be accounted for in the IBD(p) efficiency and as a potential $^{40}$K geoneutrino background if the e$^+$ signal is in accidental coincidence with another event resembling an 87 keV point energy deposition – the latter is likely to be a very rare occurrence.

As shown in Figure 5, the significance of the $^{40}$K geoneutrino signal detection, or the rejection of the null hypothesis, reaches the level of 5 (3) sigma for a detector mass slightly under 240 (90) kilotons and 50% loading of copper, with 10 years of data taking. Details are shown in Table 3. Despite the robustness of the novel detection methodology being proposed, the discovery potential for potassium geoneutrinos remains a major experimental challenge due to the small event rate of K geoneutrinos. Even in a hypothetical background-free scenario, this is largely unavoidable as it is due to the small acceptance caused by the narrow detection energy window, [1.0,1.3] MeV, where K geoneutrinos can be detected via charged-current interactions, along with the interaction cross section and the required number of target nuclei. Hence, the discovery of K geoneutrinos via this methodology, if proved feasible, would inexorably demand a *LiquidO* detector of the size of Hyper-Kamiokande [46] even in the optimistic – experimentally not demonstrated – scenario where doping is postulated at the 50% level. The *LiquidO* detector technology is a must for the unique antimatter detection signature that enables the necessary huge background suppression. The delayed coincidence with the $^{63}$Ni$^*$ deexcitation gamma ray is another strong feature adding robustness to the overall methodology.

Figure 5 illustrates two loading cases for comparison: 10% and 50%, by weight fraction. Loading at 10% represents the amount of loading that has been considered by previous liquid scintillator experiments, for example, the proposed LENS R&D programme [47], thus corresponding to the present state-of-the-art. High loading above ≥10% would require a further vigorous R&D effort. In either approach, maximising the light yield of the scintillator cocktail is important due to the low kinetic energy of the e$^+$ originating from IBD($^{63}$Cu). Indeed, the e$^+$ that is produced from $^{40}$K geoneutrino capture on $^{63}$Cu will deposit no more than ~100 keV energy. For the detection of the e$^+$, both annihilation gamma rays are readily observed and can be traced back to their common point of origin. Nevertheless, registering the signal from the energy deposited by the e$^+$ provides the baseline for the distinctive antimatter event signature in *LiquidO*. This sets the target for detector light collection and detection to be a minimum of around 200 photoelectrons/MeV, corresponding to ~20 photoelectrons for the kinetic energy resulting from a $^{40}$K geoneutrino, so as to preserve good sensitivity to these low energy events.

Despite the enormous challenges posed by this problem, the methodology and basic concepts of our proposed approach to K geoneutrino discovery is robust and have a grounded, experimental development path. Our work aimed to fully understand what it would take to pursue K geoneutrino detection using a charged-current interaction, thereby exploiting the special antimatter signature of the geoneutrino events, thus providing the framework for fully envisaging this unique and exciting potential.

# Conclusions

In this article, we proposed a novel experimental methodology where the detection of geoneutrinos exploits the $e^+$ (antimatter) tagging ability of the new *LiquidO* detection technique. We propose to use charged-current antineutrino interactions on $^{63}$Cu. $^{63}$Cu has been identified as the only known isotope today capable of meeting all the conditions necessary for enabling $^{40}$K geoneutrino detection with a robust discovery potential. A huge *LiquidO* scintillation detector of about 240 (90) kilotons mass with 50% Cu loading appears to be the minimal configuration to reach 5 (3) sigma discovery potential upon 10 years of exposure. In the same detector, the simultaneous detection of U and Th geoneutrinos using conventional IBD interactions on protons would yield the highest statistical sample ever (order 35,000 events), leading to permille statistical precision of the U and Th geoneutrino fluxes. These precise measurements, combined, would be essential for discriminating between competing models of Earth's chemical composition and the distribution of the main heat-producing elements (U, Th and K) [48]. An extra advantage of this approach is that most, if not all, of the experimental feasibility of the methodology could be established via dedicated data-driven "test-beam" measurements using a small *LiquidO* detector of a few-tons scale, located close to a nuclear reactor, similar to that foreseen in the CLOUD/AntiMatter-OTech scientific programme [49].

Our complete understanding of the Earth today and its formation would greatly benefit from the observation of $^{40}$K geoneutrinos, whose rate measurement would constitute a major discovery. The impact of the first $^{40}$K geoneutrino measurement would be a direct determination of the bulk mass of potassium in the Earth and its abundance in the deep interior after accounting for the actual knowledge of the amount of potassium in the accessible lithosphere. Current geoneutrino measurements provide radiogenic heat power estimates relying on a model-dependent K/U ratio. Instead, a direct measurement of the $^{40}$K heat power would provide an experimental constraint to the radiogenic fraction of the Earth's internal *heat budget*. For instance, a fully chondritic Earth would produce a very large $^{40}$K signal [50], hence new confirmation of such an Earth model would be possible. Moreover, measuring the K/U ratio would also provide critical information about the behaviour of volatile elements during Earth's early-stage formation [51]. The detector being proposed would likely provide the first unconstrained evaluation of the Th/U ratio in the bulk Earth, another value that tests bulk composition assumptions based on chondritic models. Finally, a direct measurement of the $^{40}$K geoneutrinos would be crucial to shed light on the "missing K and Ar" mysteries and, in turn, provide insights into the Earth's composition, structure, and thermal evolution. The K geoneutrino discovery stands as the most important quest and challenge in this field of research today.

# Acknowledgements

We acknowledge the support received from: i) the "Chaire Internationale de Recherche Blaise Pascal" (Laureate 2016: Prof F. Suekane) financed by Région Île-de-France (Paris) and coordinated by the Fondation de l'École Normale Supérieure (Paris) and hosted by CNRS at the APC Laboratory (Paris). This funding enabled launching the pioneering studies, including supporting a dedicated postdoctoral research fellow (Dr S. Wagner) whose studies, input and comments were critical; ii) "Istituto Universitario degli Studi Superiori di Ferrara 1391" for supporting Copernicus Visiting Scientist (Laureate 2019: Prof M. Chen) and hosted by the Università di Ferrara and INFN Ferrara during the key phase of this study; and iii) the "Emilie du Châtelet" Program, financed by the P2IO (LabEx, Université Paris-Saclay, France) supporting visiting scientist (2022: Prof M. Chen) hosted at the IJCLab during the final writeup phase of this publication. We also thank Prof C. Volpe (APC, Paris) for helpful discussions regarding nuclear physics input for neutrino cross sections. The authors pay tribute to Prof Giovanni Fiorentini for his pioneering studies on geoneutrinos. Although he passed away in June 2022, his memory remains vibrant in our minds.

## Author Contributions

The primary authors for this work have been listed first (alphabetically), followed by all other LiquidO consortium members.

## Competing Interest Declaration

The authors declare no competing interests.

## Additional Information

No additional information.

# Methods

## Glossary

- **Bulk Silicate Earth (BSE):** (see also primitive mantle) original chemical composition of the silicate part of the Earth after core accretion and separation and prior to crust differentiation.
- **Bulk Silicate Earth model:** compositional model of the Bulk Silicate Earth based on chemical, cosmochemical or physical dynamical assumptions and/or direct observations.
- **Carbonaceous chondrites CI:** most primitive meteorites with the highest proportion of volatile compounds and thought to have been formed the farthest from the Sun of any of the chondrites. Aside from lithium and gaseous volatile elements, they have a chemical composition close to the one measured in the solar photosphere. The presence of volatile organic chemicals and water indicates that they have not undergone significant heating (>200 $°C$).
- **Chondrites:** undifferentiated stony meteorites (i.e. meteorites not modified due to melting or differentiation of the parent body) formed from primitive asteroids resulting from the accretion of various types of dust and small grains present in the early solar system.
- **Core:** innermost metallic portion of the Earth separated (thickness ~3480 km) from the mantle by the core-mantle boundary. Seismic measurements highlight the presence of a solid inner core and a liquid outer core.
- **Crust:** outermost solid shell of the Earth (average thickness ~30 km). It comprises the lighter and older continental crust (thickness ~34 km) and the denser and younger oceanic crust (thickness ~8 km). The continental crust can be further distinguished into three different layers having comparable thicknesses (~10 km), from bottom to top: the lower crust, the middle crust and the upper crust. On top of the crust sits a thin (~1 km) layer of lighter sediments.
- **Earth differentiation:** separation of different portions of the Earth due to different physical and/or chemical affinities of the elements. The 1$^{st}$ differentiation occurred during the accretion phase and gave rise to the Earth's core, where heavy metallic siderophile elements accumulated, and to the undifferentiated mantle, i.e. the primitive mantle. The 2$^{nd}$ differentiation took place later during the cooling of the primitive mantle, creating the crust and the modern mantle, with incompatible elements (unsuitable in size or charge to the cation sites of the surrounding minerals) preferably accumulating in the crust. Convective and tectonic processes occurred after the 2$^{nd}$ differentiation stage, and still active, lead to the formation of the new crust (oceanic crust) and to the recycling of continental crust (up to 10 times).
- **Earth's heat budget:** amount of heat produced inside the Earth. Heat flow measurements over the Earth surface allow to estimate a ~47 TW heat budget resulting from the superposition of two main inputs: (i) the radiogenic heat produced by the natural occurring decays of the heat-producing elements (HPEs) and (ii) the secular cooling occurring since Earth's formation when gravitational binding energy was released due to matter accretion.
- **Earth's heat power:** see Earth's heat budget's definition.
- **Energy flow:** the topology of the energy deposition inside the scintillator as a function of time. In *LiquidO* the features contained in the spatial pattern and in the time profile of the light collected (proportional to the energy deposited) provide powerful particle identification capabilities.
- **Enstatite chondrites EH:** the most chemically reduced meteorites containing iron in the form of metal or sulphide rather than of oxide. Their composition is rather different from the Sun's, but they are the only meteorites having the same isotopic composition as terrestrial samples. Moreover, they are largely degassed and have sufficiently high iron content to explain Earth's volatilization and its metallic core.
- **Geoneutrinos:** electron antineutrinos (and neutrinos) emitted in the β decays of the long-lived naturally occurring radionuclides present in the Earth. These isotopes include $^{40}$K, $^{87}$Rb, $^{113}$Cd, $^{115}$In, $^{138}$La, $^{176}$Lu, $^{187}$Re and the elements belonging to the decay chains of $^{232}$Th, $^{235}$U and $^{238}$U. The most important in terms of geoneutrino luminosity are $^{40}$K and the ones belonging to $^{232}$Th and $^{238}$U decay chains, with only the latter two observable with present detection techniques. Different from the other mentioned isotopes (which only undergo β⁻ decays), $^{40}$K decays can emit both neutrinos and antineutrinos. However, the detection of neutrinos is prevented by their low energy and the overwhelming solar neutrino flux which is roughly three orders of magnitude higher.

- **Heat-producing elements (HPEs):** long-lived radioactive elements naturally occurring inside the Earth whose decays are producing heat since the formation of our planet up to the present day. The most important HPEs in terms of generated heat power are U, Th and K, which contribute to more than 99% of Earth's internal radiogenic power.
- **LiquidO:** a novel detection technique using an opaque scintillator and a lattice of optical fibres to confine and collect light near its creation point (i.e. the vertex of energy deposition). The usage of fibres connected to silicon photomultipliers and the tolerance to opacity enables high resolution imaging and a natural affinity for doping.
- **Lithophile elements:** elements which preferably bind with oxygen in oxides and silicates (e.g. K, Na, Ca, Mg, Al, U, Th and Ti) and in turn tend to remain on or close to the uppermost Earth layers.
- **Lithosphere:** outermost shell of the Earth comprising the crust and the continental lithospheric mantle, i.e., the uppermost, rigid part of the mantle (thickness ~170 km).
- **Mantle:** Earth's inner layer sitting between the crust and the core. It can be distinguished into separate layers according to geophysical and geochemical arguments. In particular a separation in upper mantle (UM) and lower mantle (LM) is recognized in terms of the seismic velocity discontinuity located at ~650 km depth from the surface, which translates into an average UM and LM thickness respectively of 590 km and 2220 km. Geochemical arguments define a depleted mantle (DM) with a mean thickness of 2090 km distinguished at a ~2200 km depth from the surface, lying on top of an enriched mantle (EM) with a mean thickness of 710 km.
- **Missing K:** discrepancy between the estimates of potassium amount in the Earth estimated by geochemical and by cosmochemical arguments. By comparing meteoritic abundances with Earth's estimated abundances, it turns out that the Earth retained only ~1/3 to ~1/8 of its initial potassium's mass. Two theories on the fate of the "missing K" include loss to space during accretion or segregation into the accreting core, but no experimental evidence has been able to confirm or rule out any of the hypotheses, yet.
- **Primitive mantle (PM):** portion of the Earth outside the metallic core. It corresponds to the actual lithosphere and mantle portions which were a unique chunk before the 2$^{nd}$ differentiation evolution stage occurred after the separation of the core into the solid and liquid phases (1$^{st}$ differentiation).
- **Refractory elements:** elements which tend to retain their solid form at high temperatures and pressures.
- **Secular cooling:** dissipation of the primordial heat leftover from the formation of the Earth.
- **Volatile elements:** elements which tend to occur as gases or volatile hydrides, strongly depleted on Earth compared to the Sun. If not escaping from the Earth, they remain mostly on the surface or in the atmosphere as they occur in liquids and/or gases at surface temperatures and pressures.
- **Volatility:** tendency of an element to vaporise and quantified by the vapour pressure: the higher the vapour pressure of a liquid at a given temperature, the higher the volatility.

# Geoneutrino Signal Calculation

The prediction of the IBD geoneutrino signal at a given experimental site requires the modelling of the three geoneutrino life-stages, i.e. (i) production inside the Earth, (ii) propagation to the detector site and (iii) detection via the IBD reaction on a given target.

For a given Earth elemental volume, U, Th and K activities (i.e., the average decay rates) are separately computed as the ratio between the number of radioactive nuclei and the corresponding radioisotope mean lifetime. The oscillated geoneutrino flux is obtained by weighting the activities for the corresponding geoneutrino spectrum (normalised to the number of geoneutrinos per decay) [9, 52], by scaling for the isotropic $1/4\pi r^2$ spherical factor, and by applying the electron antineutrino three-flavour survival probability [53] with up-to-date oscillation parameters [54]. Finally, the U, Th and K geoneutrino IBD signals per unit time and unit target isotope are calculated by convolving the oscillated geoneutrino spectra with the IBD cross section for the target isotope of interest. Expected signals in Terrestrial Neutrino Units (TNU) are determined by assuming a one-year acquisition time and $10^{32}$ IBD target nuclei as follows [9, 55]:

$$S_{i,n}(\vec{r}) = \frac{IA_n N_{target} \cdot T}{m_i \cdot \tau_i} \cdot \int dE_{\bar{\nu}} \cdot Sp_i(E_{\bar{\nu}}) \cdot \sigma_n(E_{\bar{\nu}}) \int d^3r' \cdot \frac{C_i \cdot a_i(\vec{r}') \cdot \rho(\vec{r}')}{4\pi |\vec{r} - \vec{r}'|^2} \cdot P_{ee}(E_{\bar{\nu}}, |\vec{r} - \vec{r}'|)$$

where $i$ runs over the *HPE* ($i$=$^{238}$U, $^{232}$Th and $^{40}$K) and $n$ runs over the target isotope (n=$^{1}$H, $^{3}$He, $^{14}$N, $^{33}$S, $^{35}$Cl, $^{45}$Sc, $^{63}$Cu, $^{79}$Br, $^{87}$Sr, $^{93}$Nb, $^{106}$Cd, $^{107}$Ag, $^{135}$Ba, $^{147}$Sm, $^{151}$Eu, $^{155}$Gd, $^{171}$Yb and $^{187}$Os), $IA_n$ is the isotopic abundance of the IBD target isotope, $N_{target}$ is equal to $10^{32}$, $T$ is a one-year acquisition time [s], $m_i$ is the atomic mass of the *HPE* [g], $\tau_i$ is the mean lifetime of the *HPE* [s], $Sp_i(E_{\bar{\nu}})$ is the geoneutrino energy spectrum for the *i-th HPE* [#$\bar{\nu}$ MeV$^{-1}$], $\sigma_n(E_{\bar{\nu}})$ is the IBD cross section for the *n-th* target isotope [cm$^2$], $C_i$ is the isotopic abundance of the *ith HPE* [g/g], $a_i(\vec{r}')$ is the mass abundance [g/g] of the *ith HPE* in the elemental volume $d^3r'$, $\rho(\vec{r}')$ is the volumetric density of the elemental volume $d^3r'$ [g/cm$^3$], $\vec{r}$ is the position of the experimental site with respect to the centre of the Earth, $\vec{r}'$ is the position of the elemental volume $d^3r'$ with respect to the centre of the Earth and $P_{ee}(E_{\bar{\nu}}, |\vec{r} - \vec{r}'|)$ is the electron antineutrino survival probability [dimensionless] for an antineutrino with energy $E_{\bar{\nu}}$ traveling for a distance $|\vec{r} - \vec{r}'|$ [cm] from the emission point in the elemental volume to the detector position.

In order to perform this geoneutrino signal calculation, it is necessary to adopt a 3-dimensional voxel-wise Earth model according to which each elemental volume is assigned with an *HPE* abundance and a volumetric density. At this scope, the Earth is typically divided into its two main HPEs-bearing reservoirs, i.e. the lithosphere (shallow and relatively rich in HPEs) and the mantle (thick and relatively poor in HPEs) [45]. The lithosphere is the outermost Earth shell with an average thickness of 170 km, comprising (from top to bottom) sediments, continental or oceanic crust and continental lithospheric mantle, while the mantle has a typical thickness of 2800 km and extends from the bottom of the lithosphere to the core-mantle boundary.

The geophysical structure of the Earth is quite well established from seismic and gravimetric measurements both in terms of reservoir thicknesses and density [45]; on the other hand, a wide range of compositional models of the Bulk Silicate Earth (BSE) has been proposed in the past decades. In particular, three are the most popular classes of BSE compositional models, typically referred to as "cosmochemical", "geochemical" and "geodynamical", which are respectively based on EH enstatite chondrites composition (low HPE abundances), CI carbonaceous chondrites composition (moderate HPE abundances) and on the energetics of mantle convection and the observed surface heat loss (high HPE abundances) [48]. Following [14], in the panorama of available compositional models, we call "cosmochemical" the model of [56], "geochemical" the model of [1] with K abundances corrected following [57] and "geodynamical" the model of [58], which provides the HPE masses in the BSE (M$_{BSE}$) reported in Table 4.

**Table 4 | Bulk Silicate Earth compositional models.** Uranium, thorium and potassium masses in the BSE (i.e. actual lithosphere plus mantle reservoirs) for the "cosmochemical", "geochemical" and "geodynamical" models adopted in the present study and taken from [14]. HPE masses for the BSE are calculated multiplying each HPE expected abundance by the BSE bulk mass of $4.035 \cdot 10^{24}$ kg. The "cosmochemical" BSE abundances are taken from [48] and based on the enstatite BSE model of [56]. The "geochemical" BSE abundances are taken from [1] with some minor corrections for the K abundances taken after [51]. Finally, the "geodynamical" BSE abundances are taken from [48], which is based on [59] estimates.

|               | $M_{BSE}(U)$ [$10^{16}$kg] | $M_{BSE}(Th)$ [$10^{16}$kg] | $M_{BSE}(K)$ [$10^{19}$kg] |
|---------------|---------------------------|-----------------------------|----------------------------|
| Cosmochemical | 5 ± 1                     | 17 ± 2                      | 59 ± 12                    |
| Geochemical   | 8 ± 2                     | 32 ± 5                      | 113 ± 24                   |
| Geodynamical  | 14 ± 2                    | 57 ± 6                      | 142 ± 14                   |

In the context of possible HPE distributions in the Earth, we rely on the fact that for each HPE, the average mass in the lithosphere ($M_{litho}$), together with its standard deviations ($\sigma_{litho}$), is statistically well known from direct measurements on rock samples [45]. In this perspective, subtracting $M_{litho}$ from the $M_{BSE}$, the residual HPE masses are assigned to the mantle ($M_{mantle}=M_{BSE}-M_{litho}$) according to 3 different possible distributions: (i) in a 10 km thick layer at the core-mantle boundary [60], (ii) in a 710 km thick enriched mantle layer sitting at the core-mantle boundary and underlying a 2090 km thick depleted mantle [45], (iii) in a 2800 km thick homogeneous mantle [61]. The core is considered devoid of HPEs.

Across the possible combinations of HPE masses and distributions, a "low", "medium" and "high" scenario were built by taking into account the following points: i) the geophysical structure of the reservoirs belonging to the lithosphere is fixed and taken after [14]; ii) the lithospheric HPE masses can vary in the $\sigma_{litho}$ range, iii) the "proximity argument" holds [9], which states that "the minimal (maximal) contributed flux is obtained by placing HPE as far (close) as possible to the detector". Figure 6 and Table 5 provide, respectively, a visual sketch and a quantitative description of the relevant features of the "low", "medium" and "high" scenarios that give rise to, respectively, the minimum, central and maximum geoneutrino fluxes and signals (see also Figure 1 and Table 1), defining the expected signal and its variability range. The resulting HPEs masses adopted for the signal calculation according to the three different scenarios are reported in Table 6.

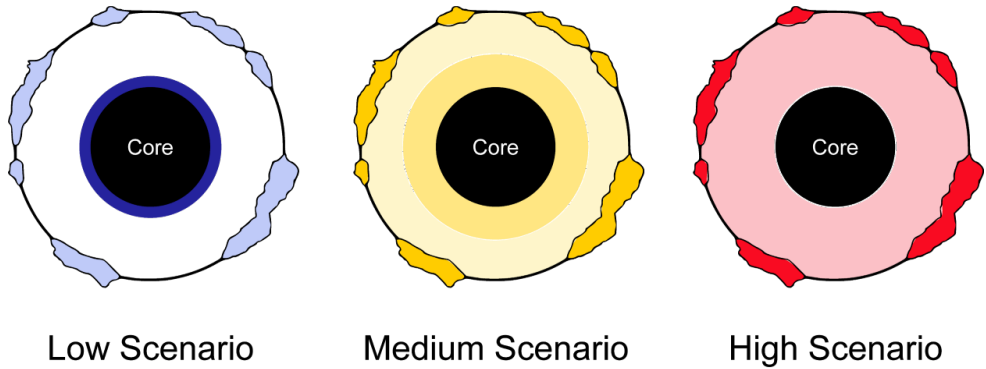

**Figure 6 | Different scenarios for the amount and distribution of HPEs in the BSE.** Visual sketch of the "low", "medium" and "high" Earth scenarios, respectively providing the minimum, central and maximum geoneutrino fluxes and signals for a given experimental site location. Portions of the Earth filled with colour describe HPE bearing reservoirs, where the colour intensity is proportional to the HPE abundance. Blue, yellow and red colours indicate progressively higher HPE masses (see also Table 4 and Table 5).

Table 5 | Rationale adopted for assigning the Heat Producing Element masses in the BSE, lithosphere and mantle. "Low", "medium" and "high" scenarios are considered for the evaluation of the geoneutrino signal variability range (see also Figure 6). For each HPE, the masses in the Bulk Silicate Earth ($M_{BSE}$) for the "cosmochemical" (cosmo), "geochemical" (geoch) and "geodynamical" (geodyn) models are the ones reported in Table 4. For each HPE, the lithospheric mass ($M_{litho}$) and its standard deviation ($1\sigma_{litho}$) are taken after [45]. For the medium Scenario, the HPE masses in the depleted and enriched mantle are determined on the basis of a mass balance argument as reported in [45], where $f_{mantle}$ corresponds to the mantle mass fraction of the specific HPE in the depleted mantle ($f_{mantle}(U)=0.52$, $f_{mantle}(Th)=0.38$, $f_{mantle}(K)=0.63$).

|  | Low scenario | Medium scenario | High scenario |
|---|---|---|---|
| $M_{BSE}$ | $M_{BSE}^{cosmo}$ | $M_{BSE}^{geoch}$ | $M_{BSE}^{geodyn}$ |
| $M_{litho}$ | $M_{litho}-1\sigma_{litho}$ | $M_{litho}$ | $M_{litho}+1\sigma_{litho}$ |
| $M_{mantle}$ | $M_{BSE}^{cosmo}-(M_{litho}-1\sigma_{litho})$ | $F_{mantle}\cdot(M_{BSE}^{geoch}-M_{litho})$ <br> $(1-f_{mantle})\cdot(M_{BSE}^{geoch}-M_{litho})$ | $M_{BSE}^{geodyn}-(M_{litho}+1\sigma_{litho})$ |

Table 6 | Heat Producing Element masses in the BSE, lithosphere and mantle. For each HPE, the mantle mass ($M_{mantle}$) is obtained subtracting the lithospheric mass ($M_{litho}$), taken from [14] from the masses in the Bulk Silicate Earth ($M_{BSE}$) (Table 4). For the low scenario, the $M_{mantle}$ refers to the masses in the thin (10 km) layer above the core-mantle boundary; for the medium scenario the $M_{mantle}$ in the depleted (upper row) and enriched mantle (lower row) are separately reported; for the high scenario the $M_{mantle}$ is homogeneously distributed.

|  | Low scenario | | | Medium scenario | | | High scenario | | |
|---|---|---|---|---|---|---|---|---|---|
|  | M(U) [$10^{16}$kg] | M(Th) [$10^{16}$kg] | M(K) [$10^{19}$kg] | M(U) [$10^{16}$kg] | M(Th) [$10^{16}$kg] | M(K) [$10^{19}$kg] | M(U) [$10^{16}$kg] | M(Th) [$10^{16}$kg] | M(K) [$10^{19}$kg] |
| *BSE* | 5 | 17 | 59 | 8 | 32 | 113 | 14 | 57 | 142 |
| *Lithosphere* | 2.7 | 11.5 | 30.9 | 3.3 | 14.3 | 36.9 | 4.1 | 19.1 | 45.3 |
| *Mantle* | 2.3 | 5.5 | 28.1 | 2.44 | 6.73 | 47.9 | 9.9 | 37.9 | 96.7 |
|  |  |  |  | 2.26 | 11.0 | 28.2 |  |  |  |

The above-mentioned methodology is followed in this work to calculate the geoneutrino signals, and their variability ranges, on different IBD targets expected at Laboratori Nazionali del Gran Sasso (LNGS; 42.45° N, 13.57° E), as pertinent example site in this publication where much information exists. For the sake of completeness, we underline that the intensity of the signal strongly depends on the experimental site position. Provided that the mantle contribution is isotropic, the following arguments need to be considered: i) although the average crustal thickness is about 35 km, it can range from ~5 km (thinnest oceanic crust) up to ~70 km (thickest continental crust); ii) the continental crust is richer in HPE (~40 higher mass abundances) compared to the oceanic crust. In this framework, LNGS, sitting on top of ~35 km continental crust, represents an intermediate case. Two extreme examples of "oceanic" and "continental" geoneutrino potential experimental sites could be represented by Hawaii (19.72 N, 156.32 W) and the Himalayas (33.00 N, 85.00 E), respectively. Indeed, according to the "medium" scenario, at a Himalayan site, the geoneutrino signal ($S_{IBD(p)}$ (U+Th) = 58 TNU) is expected to be generated by HPE distributed in the lithosphere with a fraction of about 85%; while at Hawaii, approximately 75% of the geoneutrino signal ($S_{IBD(p)}$ (U+Th) = 12 TNU) originates from the mantle [45].

## Reactor Antineutrino Signal Calculation

The expected reactor signal at LNGS, as example site, was estimated as the superposition of the signals generated by all commercial nuclear power plants [43], where the parameterised antineutrino spectral shape [62] per decay was adopted to extrapolate the reactor spectrum down to the $^{40}$K geoneutrino energy window. In this respect, LNGS represents an experimental location with a long "baseline", as the closest commercial reactor is at ~400 km distance and produces the largest signal fraction (only ~3% of the total reactor signal). The annual expected reactor neutrino signal is almost constant (as each distant reactor provides only a small fraction of the signal, hence variations due to any given core's operation are not so significant) and provides about 35% of the IBD(p) signal in the geoneutrino energy window [1.806 ; 3.272] MeV and a small (~1%) fraction of the IBD($^{63}$Cu) signal in the $^{40}$K geoneutrino energy window [1.176 ; 1.311] MeV.

## Inverse Beta Decay Cross Section Calculation and Expected Geoneutrino Spectra

The nuclear matrix element for an antineutrino capture reaction $\bar{\nu}_e + {}^{A}_{Z}X \rightarrow e^+ + {}^{A}_{Z-1}Y$ is similar to the one from the related β decay ${}^{A}_{Z-1}Y \rightarrow {}^{A}_{Z}X + e^- + \bar{\nu}_e$ and can be derived from its *ft* value, or comparative half-life, which typically provides a useful criterion for the classification of radioactive transitions as allowed or forbidden to various degrees. The antineutrino capture reaction has a target dependent energy threshold $E_{th} = Q_\beta + 2m_e$, where $Q_\beta$ is the total energy released in ${}^{A}_{Z-1}Y$ nuclear β decay. The total charged current cross section (in natural units) can be calculated directly from evaluating the appropriate β-decay reaction and correcting for the spin of the system as follows [63]:

$$\sigma = \frac{2\pi^2 \ln 2}{ft \, m_e^5} \, p_e E_e \, F(E_e, Z_f) \frac{(2 J_f + 1)}{(2 J_i + 1)}$$

where *ft* is the comparative half-life of the β decaying nucleus ${}^{A}_{Z-1}Y$, taken from the ENDSF nuclear database [25], $m_e$, $p_e$ and $E_e = E_{\bar{\nu}_e} - E_{th} + m_e$ are, respectively, the e$^+$ mass, momentum and total energy, $F(E_e, Z_f)$ is the Fermi nuclear function correcting for Coulomb repulsion between the e$^+$ and the nucleus of the final state ${}^{A}_{Z-1}Y$ having charge $Z_f = Z - 1$, $\frac{(2 J_f + 1)}{(2 J_i + 1)}$ is the spin correction factor for a final state ${}^{A}_{Z-1}Y$ with spin $J_f$ and initial state ${}^{A}_{Z}X$ with spin $J_i$.

In view of $^{40}$K geoneutrino detection via the antineutrino capture reaction, the target isotope candidates were selected considering the following constraints: i) $E_{th}$ < 1.311 MeV, and ii) relatively low *ft* values (i.e. relatively high cross section). The Nuclear Data Section of the IAEA provides a user-friendly API[5] that can be used to query nuclear data and investigate isotope properties. According to the Evaluated Nuclear Structure Data File (ENSDF), contained in the IAEA database [25], a total of 515 different β$^-$ emitters exist for a total of 523 possible transitions to ground state. We are interested in these β$^-$ decaying isotopes since their products are suitable targets for the inverse process, namely the IBD. Among these 515 emitters, we are interested in those having a threshold at $E_{th}$<1.311 MeV, the energy endpoint of the $^{40}$K geoneutrino spectrum. There are a total of 26 possible targets satisfying this condition, not all, however, stable. Since a realistic geoneutrino experiment will require a large number of target nuclei and a high level of radiopurity, this consideration is sufficient of itself alone to discard all the unstable candidates. We thus discuss only targets which are stable or can be considered stable on the time scale of a geoneutrino experiment (as the ones having half-lives comparable or higher than

---

[5] Instructions available at https://www-nds.iaea.org/relnsd/vcharthtml/api_v0_guide.html

Earth's age, $^{151}$Eu and $^{147}$Sm), ending up with a total of 17 targets leading to 20 possible IBD transitions. This led to the identification of 17 target isotopes: $^{3}$He, $^{14}$N, $^{33}$S, $^{35}$Cl, $^{45}$Sc, $^{63}$Cu, $^{79}$Br, $^{87}$Sr, $^{93}$Nb, $^{106}$Cd, $^{107}$Ag, $^{135}$Ba, $^{147}$Sm, $^{151}$Eu, $^{155}$Gd, $^{171}$Yb and $^{187}$Os, shown in Table 7, and Figure 7.

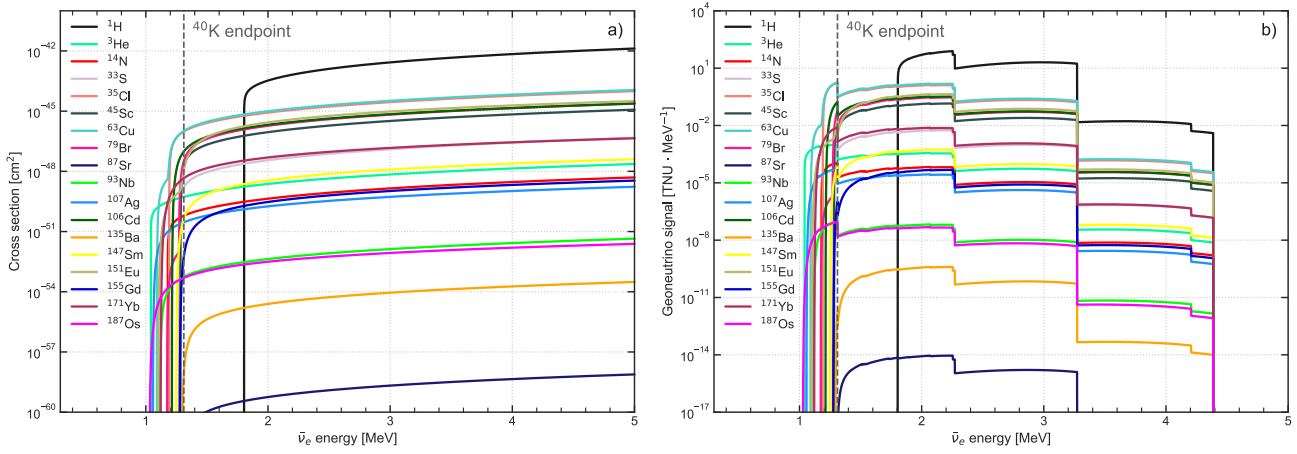

**Figure 7 | Cross sections and expected geoneutrino measured spectra for the $^{1}$H, $^{3}$He, $^{14}$N, $^{33}$S, $^{35}$Cl, $^{45}$Sc, $^{63}$Cu, $^{79}$Br, $^{87}$Sr, $^{93}$Nb, $^{106}$Cd, $^{107}$Ag, $^{135}$Ba, $^{147}$Sm, $^{151}$Eu, $^{155}$Gd, $^{171}$Yb and $^{187}$Os Inverse Beta Decay (IBD) targets.** Plot **a)** shows the cross section for the IBD reaction for the 18 different targets as a function of the incoming antineutrino energy, weighted by the corresponding isotopic abundance of each target (see Table 1). Plot **b)** shows the expected geoneutrino measured spectra at the Laboratori Nazionali del Gran Sasso (Italy) originating from uranium, thorium and potassium distributed in the Earth's lithosphere and mantle as described in [45]. The expected geoneutrino spectra are given in Terrestrial Neutrino Units (TNUs), corresponding to a 1-year acquisition time and $10^{32}$ atoms for each chemical species. In both panels the grey dashed vertical line indicates the endpoint of the potassium geoneutrino spectrum at 1.311 MeV.

Candidates with low $E_{th}$ values and with possible allowed transitions to an excited state are $^{63}$Cu, $^{79}$Br and $^{151}$Eu. The $\bar{\nu}_e$ reaction energy threshold must include the energy of the excited state and still be below the $^{40}$K geoneutrino spectrum endpoint; that is the case for all three of these isotopes and their possible (allowed) excited states. However, only an approximate value of Log($ft$)=5 [22] is assumed for the excited states, with an admissible range of 4-6 (with some even larger values occurring) [64]. Indeed, while for $^{79}$Br → $^{79}$Se* the Log($ft$) value can be analytically estimated (using the $^{79}$Se* → $^{79}$Br branching ratio, its half-life and the Fermi functions of the two states), for $^{63}$Cu → $^{63}$Ni* and $^{151}$Eu → $^{151}$Sm* the lack of knowledge of the branching ratio and/or the half-life of the final state prohibits a Log($ft$) calculation. As the cross section is inversely proportional to $ft$, a 2-unit change in Log($ft$) will result in a 2 orders of magnitude variation in the cross section. Even a small (10%) increase (decrease) in Log($ft$) will provide a cross section that is reduced (enhanced) by a factor 3 with respect to the central value estimate. This large uncertainty calls for refined nuclear physics input coming from theory and/or experiments, in particular for $^{63}$Cu (see Figure 7), which is the most promising candidate. Measurements using nuclear reactor neutrinos, as a source, will enable a direct measure of the $^{63}$Cu ability for geoneutrino detection relative to the IBD(p) interaction, thus, bypassing other complex nuclear physics effects that may modify the first order rate amplitude given by Log($ft$).

$^{3}$He would be an excellent target with a high cross section (low $ft$) and one of the lowest energy thresholds. However, its scarcity (1.3·10$^{-4}$ % isotopic abundance) and extremely high cost would make its choice prohibitive. Similar arguments apply to $^{106}$Cd, $^{33}$S, which would require isotopic enrichment to increase isotopic abundance from the natural 1.25% and 0.75%. Moreover, only an upper limit for the $^{106}$Ag → $^{106}$Cd beta decay branching ratio is available (BR<1% [25]), limiting the knowledge of $^{106}$Cd cross section to an estimated maximum value (Log($ft$) > 4.1). For $^{45}$Sc, $^{79}$Br, $^{87}$Sr, $^{135}$Ba, $^{147}$Sm, $^{151}$Eu and $^{155}$Gd the energy threshold is too close to the $^{40}$K endpoint to enable the detection of a relevant portion of the spectrum (Table 7).

**Table 7 | Inverse Beta Decay (IBD) target isotopes and expected geoneutrino signals.** The 1st column lists the target and product atoms involved in the IBD reaction, the 2nd column the IBD target Isotopic Abundance (IA) in percentage, the 3rd column the IBD reaction energy threshold ($E_{th}$) in MeV, the 4th column the Log(ft) value for the corresponding β decay of the final state. The IA, $E_{th}$ and Log(ft) values are taken by the ENSDF database [25] which points to the literature references disclosed in the 5th column. Column 6th, 7th and 8th report as central values the expected geoneutrino signal in the [$E_{th}$, 3.272] MeV energy range at LNGS originating respectively from uranium, thorium and potassium distributed in the Earth's lithosphere and mantle according to the "medium" scenario. The range in square brackets provides a variability on the expected signal in terms of minimum and maximum values obtained by respectively adopting a "low" and "high" scenario for the masses and distributions of heat-producing elements in the Earth determined with different BSE models. The expected geoneutrino signals are given in Terrestrial Neutrino Units (TNU), corresponding to a 1-year acquisition time and $10^{32}$ atoms for each chemical species (i.e. a number of IBD target atoms corresponding to $10^{32}$ scaled by the isotopic abundance).

| Target process | IA [%] | $E_{th}$ [MeV] | Log(ft) | Ref | S(U) [TNU] | S(Th) [TNU] | S(K) [TNU] |
|---|---|---|---|---|---|---|---|
| $^1H \to ^1n$ | 99.99 | 1.806 | 3.0170 | [26] | 31.5 [24.0 ; 47.0] | 9.0 [6.4 ; 14.1] | / |
| $^{63}Cu \to ^{63}Ni$ | 69.15 | 1.089 | 6.7 | [25] | 0.85 [0.64 ; 1.26] | 0.49 [0.35 ; 0.77] | 0.10 [0.07 ; 0.13] |
| $^{63}Cu \to ^{63}Ni*$ | | 1.176 | 5 | [22] | | | |
| $^{35}Cl \to ^{35}S$ | 75.76 | 1.189 | 5.0088 | [27] | 0.73 [0.56 ; 1.09] | 0.43 [0.30 ; 0.67] | 0.10 [0.07 ; 0.13] |
| $^{106}Cd \to ^{106}Ag$ | 1.25 | 1.212 | 4.1 | [28] | $(1.7 [1.3 ; 2.6]) \cdot 10^{-1}$ | $(9.7 [6.9 ; 15.2]) \cdot 10^{-2}$ | $(5.1 [3.7 ; 6.6]) \cdot 10^{-3}$ |
| $^{79}Br \to ^{79}Se$ | 50.69 | 1.173 | 10.77 | [65] | 0.15 [0.12 ; 0.23] | $(8.3 [5.9 ; 13.0]) \cdot 10^{-2}$ | $(6.5 [4.7 ; 8.4]) \cdot 10^{-4}$ |
| $^{79}Br \to ^{79}Se*$ | | 1.268 | 5 | [22] | | | |
| $^{171}Yb \to ^{171}Tm$ | 14.09 | 1.119 | 6.318 | [66] | $(4.1 [3.1, 6.1]) \cdot 10^{-3}$ | $(2.4 [1.7, 3.8]) \cdot 10^{-3}$ | $(5.0 [3.7, 6.5]) \cdot 10^{-4}$ |
| $^{151}Eu \to ^{151}Sm$ | 47.81 | 1.099 | 7.51 | [67] | 0.22 [0.17 ; 0.33] | 0.12 [0.08 ; 0.18] | $(3.2 [2.3 ; 4.2]) \cdot 10^{-4}$ |
| $^{151}Eu \to ^{151}Sm*$ | | 1.266 | 5 | [22] | | | |
| $^{45}Sc \to ^{45}Ca$ | 100 | 1.282 | 6 | [68] | $(7.5 [5.7, 11.2]) \cdot 10^{-2}$ | $(4.1 [2.9, 6.3]) \cdot 10^{-2}$ | $(3.3 [2.4; 4.2]) \cdot 10^{-4}$ |
| $^3He \to ^3H$ | $1.34 \cdot 10^{-4}$ | 1.041 | 3.0524 | [69] | $(2.2 [1.7 ; 3.4]) \cdot 10^{-4}$ | $(1.4 [1.0 ; 2.2]) \cdot 10^{-4}$ | $(1.7 [1.2 ; 2.1]) \cdot 10^{-4}$ |
| $^{33}S \to ^{33}P$ | 0.75 | 1.271 | 5.022 | [70] | $(3.0 [2.3 ; 4.5]) \cdot 10^{-3}$ | $(1.7 [1.2 ; 2.6]) \cdot 10^{-3}$ | $(4.4 [3.2 ; 5.7]) \cdot 10^{-5}$ |
| $^{14}N \to ^{14}C$ | 99.64 | 1.178 | 9.040 | [71] | $(3.8 [2.9 ; 5.7]) \cdot 10^{-5}$ | $(2.3 [1.6 ; 3.5]) \cdot 10^{-5}$ | $(8.0 [5.8 ; 10.3]) \cdot 10^{-6}$ |
| $^{107}Ag \to ^{107}Pd$ | 51.839 | 1.056 | 9.9 | [72] | $(1.6 [1.2 ; 2.4]) \cdot 10^{-5}$ | $(9.8 [6.9 ; 15.2]) \cdot 10^{-5}$ | $(5.7 [4.1 ; 7.3]) \cdot 10^{-6}$ |
| $^{147}Sm \to ^{147}Pm$ | 14.99 | 1.246 | 7.4 | [73] | $(2.9 [2.2 ; 4.3]) \cdot 10^{-4}$ | $(1.6 [1.1 ; 2.4]) \cdot 10^{-4}$ | $(1.1 [0.8 ; 1.4]) \cdot 10^{-6}$ |
| $^{187}Os \to ^{187}Re$ | 1.96 | 1.024 | 11.195 | [74] | $(2.7 [2.1 ; 4.1]) \cdot 10^{-8}$ | $(1.7 [1.2 ; 2.7]) \cdot 10^{-8}$ | $(9.3 [6.8 ; 12.0]) \cdot 10^{-8}$ |
| $^{93}Nb \to ^{93}Zr$ | 100 | 1.113 | 12.1 | [75] | $(3.7 [2.8 ; 5.5]) \cdot 10^{-8}$ | $(2.2 [1.6 ; 3.4]) \cdot 10^{-8}$ | $(8.2 [5.9 ; 10.5]) \cdot 10^{-8}$ |
| $^{155}Gd \to ^{155}Eu$ | 14.80 | 1.274 | 8.62 | [76] | $(2.4 [1.8 ; 3.6]) \cdot 10^{-5}$ | $(1.3 [0.9 ; 1.9]) \cdot 10^{-5}$ | $(6.7 [4.9 ; 8.7]) \cdot 10^{-9}$ |
| $^{135}Ba \to ^{135}Cs$ | 6.592 | 1.291 | 13.48 | [77] | $(2.0 [1.5 ; 3.0]) \cdot 10^{-10}$ | $(1.0 [0.7 ; 1.6]) \cdot 10^{-10}$ | $(5.2 [3.8 ; 6.7]) \cdot 10^{-15}$ |
| $^{87}Sr \to ^{87}Rb$ | 7.00 | 1.304 | 17.514 | [78] | $(4.6 [3.5 ; 6.9]) \cdot 10^{-15}$ | $(2.4 [1.7 ; 3.7]) \cdot 10^{-15}$ | $(3.9 [2.9 ; 5.1]) \cdot 10^{-21}$ |

Although $^{14}$N is characterized by low Z (i.e., small Fermi Coulomb correction) and by a ~100% isotopic abundance, the $^{14}_{6}C \rightarrow {}^{14}_{7}N + e^{-} + \bar{\nu}_e$ allowed β decay is disfavoured (high *ft* value). This makes $^{14}$N a poor target choice due to its small antineutrino cross section. A similar argument applies to $^{93}$Nb, $^{171}$Yb, $^{107}$Ag and $^{187}$Os, whose poor *ft* values strongly limit the amplitude of the IBD cross section, excluding them as possible targets. In conclusion, $^{35}$Cl and $^{63}$Cu were identified as the most promising antineutrino capture target for the detection of $^{40}$K geoneutrinos.

## Comparison to Other $^{40}$K Detection Concepts

Two proposals for $^{40}$K geoneutrino detection [17, 18] suggest using the antineutrino-electron elastic scattering reaction. The advantages compared to the charged-current interaction described in this study would be i) a possible lower energy threshold, set by the detector calorimetric capabilities rather than by the interaction kinematics, ii) a cross section and a number of targets higher than most antineutrino capture reactions, and iii) the possible exploitation of directionality, if the detectors allowed. The $^{40}$K geoneutrino signal, for example, at the Laboratori Nazionali del Gran Sasso, would be 1.9 TNU, obtained by integrating the event rate for electron recoils above 800 keV [17]. Electron scattering proposals rely heavily on directionality, using the recoil electron track direction in a gaseous detector or proposing (under R&D) to reconstruct the electron's direction from the Cherenkov light produced amongst more copious scintillation light, in a denser liquid scintillator detector [18]. The recoil electron's direction is needed to distinguish $^{40}$K geoneutrinos from solar neutrinos scattering off electrons, wherein the latter has a larger flux by three orders of magnitude. Even if the solar neutrino background could be reduced by directionality, other single electron backgrounds (e.g. natural beta minus decays) would be isotropic and abundant. Techniques relying on electron detection therefore imply an extraordinary level of radiopurity of the detector better than today's extraordinary levels achieved by Borexino [14]. In comparison, this publication presents the potential of IBD($^{63}$Cu) detection for $^{40}$K geoneutrinos without directional information, which is not required because the abundant solar neutrino flux does not pose a problem.

Additionally, despite the lower event rate, IBD($^{63}$Cu→$^{63}$Ni$^*$) would generate distinctive e$^+$ signals in a *LiquidO* detector in coincidence with a delayed 87 keV gamma ray, and will have far fewer possible backgrounds compared to recoil electrons, as discussed in the main text. Clear e$^+$ identification in *LiquidO* would then exploit the unique antimatter signature and lead to higher signal-to-background ratios with much less radiopurity control required on the detector side, providing a promising experimental framework for discovery.